\documentclass[
twocolumn,
]{aastex7}
\usepackage{amsmath,amssymb}
\usepackage{ulem}
\usepackage[english]{babel}
\usepackage{graphicx} 
\usepackage{dcolumn} 
\usepackage{bm} 
\usepackage{makecell} 
\usepackage{threeparttable}
\usepackage{multirow} 
\usepackage{booktabs}
\usepackage{color}
\newcommand{\ksM}{{\rm km\,s^{-1}\,Mpc^{-1}}}

\begin{document}\begin{sloppypar}

\title{
The Use of Binary Black Holes Merging in AGN Disks for Hubble Constant Measurements
}

\author{Guo-Peng Li}
\affiliation{
Department of Astronomy, School of Physics and Technology, Wuhan University, Wuhan 430072, China
}\email{li_g.p@whu.edu.cn}

\author{Xi-Long Fan}
\affiliation{
Department of Astronomy, School of Physics and Technology, Wuhan University, Wuhan 430072, China
}\email[show]{xilong.fan@whu.edu.cn}

\correspondingauthor{Xi-Long Fan}

\date{\today}

\begin{abstract}
We study the impact of environmental effects on the measurement of the Hubble constant ($H_0$) from gravitational wave (GW) observations of binary black hole mergers residing in active galactic nuclei (AGNs) near the central supermassive black hole. Using the potential hierarchical triple merger candidate GW190514-GW190521 in AGN J124942.3+344929 with its potential electromagnetic counterpart ZTF19abanrhr as a multimessenger case study, we demonstrate that environmental effects can be negligible for mergers at approximately tens to hundreds of Schwarzschild radii from the supermassive black hole. We find $H_0=40.9_{-8.9}^{+19.3}\,{\rm km\,s^{-1}\,Mpc^{-1}}$ (median and 68\% credible interval) under a flat prior and flat $\Lambda$CDM cosmology. Incorporating GW170817 prior information improves constraints to $H_0=68.8_{-6.0}^{+7.7}\,{\rm km\,s^{-1}\,Mpc^{-1}}$. We suggest that in general, AGN environments could serve as viable laboratories for cosmological studies from GW observations where environmental effects remain below detection thresholds.
\end{abstract}

\section{Introduction}
The binary compact star mergers are promising sources for the standard siren cosmological measurements~\citep{1986Natur.323..310S,2010CQGra..27u5006S}, providing a new approach to measure the Hubble constant since gravitational wave (GW) signals provide direct measurements of luminosity distances. This approach relies on the identification of electromagnetic (EM) counterparts and/or statistical analysis of galaxies within the GW localization volume~(e.g.,~\citealp{1986Natur.323..310S,2005ApJ...629...15H,2012PhRvD..86d3011D,2013arXiv1307.2638N,2014ApJ...795...43F,2017PhRvL.119r1102F,2017Natur.551...85A,2018Natur.562..545C,2019ApJ...871L..13F,2021ApJ...909..218A,2020PhRvD.101l2001G,2023ChPhC..47f5104J,2024SCPMA..6720412J,2024ApJ...976..153L,2024PhRvL.133z1001M,2025arXiv250210780T,2025arXiv250301997C}), differing from traditional probes based on such as standard candles~\citep{2016ApJ...826...56R,2019ApJ...876...85R}, the cosmic microwave background~\citep{2016A&A...594A..13P,2020A&A...641A...6P}, and other methods~\citep{2019ApJ...882...34F,2019MNRAS.486.2184M,2019ApJ...886...61Y,2020ApJ...891L...1P}. This methodology is crucial for resolving the Hubble tension and elucidating the nature of dark energy~(e.g.,~\citealp{2020PhRvL.124p1301S,2020PhRvD.102f3527N,2021CQGra..38o3001D,2023ARNPS..73..153K,2025arXiv250211568L}). 
A landmark demonstration of Hubble constant measurement~\citep{2017Natur.551...85A} through GW signals is the binary neutron star merger GW170817~\citep{2017PhRvL.119p1101A} detected by Advanced LIGO~\citep{2015CQGra..32g4001L} and Advanced Virgo~\citep{2015CQGra..32b4001A}. The association with EM counterpart GRB 170817A enables the first GW standard siren measurement of the Hubble constant to be about $70\,\ksM$~\citep{2017Natur.551...85A}. With additional detections by LIGO, Virgo, and KAGRA~\citep{2012CQGra..29l4007S,2013PhRvD..88d3007A} are expected to improve the precision of Hubble constant constraints~\citep{2018Natur.562..545C}.

The binary black hole merger GW190521 with a total mass of around $150\,M_{\odot}$~\citep{2020PhRvL.125j1102A}, observed during the first part of the third LIGO-Virgo observing run~\citep{2021PhRvX..11b1053A}, represents an intermediate mass black hole system. Such massive systems likely originate from dynamical formation channels in dense star clusters or active galactic nuclei (AGNs,~e.g.,~\citealp{2019ApJ...876..122Y,2020ApJ...900L..13A,2020ApJ...902L..26F,2022A&A...666A.194L,2022PhRvD.105f3006L,2023PhRvD.107f3007L,2023NatAs...7...11G,2025ApJ...981..177L}). AGN disks particularly enhance massive black hole formation through orbital alignment, accretion, and hierarchical growth mechanisms~(e.g.,~\citealp{2019PhRvL.123r1101Y,2019ApJ...876..122Y,2022PhRvD.105f3006L,2023PhRvD.107f3007L,2020ApJ...898...25T,2020MNRAS.494.1203M,2024arXiv241018815D}). 
Notably, AGN environments may associate binary black hole mergers with EM emissions and/or high-energy neutrinos~(e.g.,~\citealp{2019ApJ...884L..50M,2021ApJ...916L..17W,2023ApJ...955...23T,2023MNRAS.524.6015L,2023ApJ...951...74Z,2024ApJ...961..206C,2024MNRAS.528L..88Z,2024arXiv240610904Z}). The potential EM counterpart to GW190521, an optical flare ZTF19abanrhr detected by ZTF from AGN J124942.3+344929 at $z=0.438$~(\citealp{2020PhRvL.124y1102G}, also see~\citealp{2021arXiv211212481C,2023PhRvD.108l3039M}), has motivated Hubble parameter measurements under different physical assumptions and cosmological priors~\citep{2020arXiv200914199M,2021arXiv211212481C,2021ApJ...908L..34G,2022MNRAS.513.2152C,2023PhRvD.108l3039M}.

However, environmental effects near the central supermassive black hole in AGNs may bias GW-derived luminosity distance measurements~\citep{2023PhRvD.107d3027T}. Specifically, these effects affect the merger and ringdown part of the GW signal~\citep{2025PhRvL.134h1402S}, potentially introducing systematic deviations in Hubble constant determination. As analyzed by ~\citet{2023PhRvD.108l3039M}, two primary mechanisms emerge: (1) gravitational redshift from the supermassive black hole potential, and (2) relativistic redshift induced by binary orbital motion as it orbits the supermassive black hole.

In addition, the massive component masses ($\sim$$85\,M_{\odot}$ and $\sim$$66\,M_{\odot}$) in GW190521 suggest possible hierarchical merger origins~\citep{2016ApJ...824L..12O,2017PhRvD..95l4046G,2017ApJ...840L..24F,2020ApJ...900L..13A,2020ApJ...902L..26F,2021NatAs...5..749G,2021ApJ...915L..35K,2021MNRAS.502.2049L,2024arXiv241102778C,2024PhRvL.133e1401L,2024ApJ...975..117M,2025ApJ...981..177L}. A plausible hierarchical triple merger scenario~\citep{2018MNRAS.476.1548S,2019MNRAS.482...30S} posits GW190514 as the precursor merger to GW190521~\citep{2020MNRAS.498L..46V,2021ApJ...907L..48V}, with one of both potentially associated with ZTF19abanrhr~\citep{2020PhRvL.124y1102G,2023ApJ...942...99G}. 
Recently, \citet{PhysRevD.111.103016} investigated the potential association by taking into account of sky position, distance, and mass of the sources using a Bayesian criterion. They found the association is favored over a random coincidence, with a log Bayes factor of 16.8, corresponding to an odds ratio of $\sim$$199:1$, assuming an astrophysical prior odds of $10^{-5}$.
Notably, when accounting for the primary masses of the two GW events as potential products of mergers in the AGN formation channel, the Bayes factor increases significantly, further enhancing the preference for this association by a factor of $\sim$$10^2$. This provides strong evidence for the first hierarchical triple merger with an EM counterpart in the AGN formation channel.

In this study, we investigate environmental impacts on Hubble constant measurements from AGN-embedded GW signals. Using the GW190514-GW190521 hierarchical merger pair and EM counterpart ZTF19abanrhr in AGN J124942.3+344929 as a case study, we quantify these environmental effects.

The rest of this paper is structured as follows. 
Section~\ref{sec:methods} details our methodology, including Hubble constant computation in Section~\ref{sec:HCM} and environmental effect analysis in Section~\ref{sec:EE}.
Results and discussion appear in Section~\ref{sec:RD}, followed by conclusions in Section~\ref{sec:conslusions}.

\section{Methods}\label{sec:methods}

\subsection{Hubble constant computation}\label{sec:HCM}

The probability density of the Hubble constant ($H_0$) from the GW signal ($d_{\rm GW}$) and EM signal ($d_{\rm EM}$) of a multimessenger source reads~\citep{2019ApJ...871L..13F,2021ApJ...908L..34G}
\begin{equation}\begin{aligned}
    p(H_0|d_{\rm GW},d_{\rm EM}) 
    \propto &
    \frac{p(H_0)}{\beta(H_0)}\,
    \int
    p(d_{\rm GW}|{\hat D_L}(z_{\rm EM},H_0),\Omega_{\rm EM})\,\\&\times
    p(D_L)\,p(\Omega)\,
    {\rm d}D_L\,{\rm d}\Omega \,,
    \label{eq:1}
\end{aligned}\end{equation}
where $p(H_0)$ is the prior on $H_0$; 
$\beta(H_0)$ is a normalization term that is the fraction of sources detectable at given $H_0$.
The next term is the probability of the GW data in the presence of signal at the three-dimensional spatial position of the EM counterpart.
$p(D_L)\propto D_L^2$ is the luminosity distance prior 
and $p(\Omega)$ is the sky position prior that is uniform in the sky localization.

The probability density of $H_0$ from a hierarchical triple merger pair with its EM counterpart, is to multiply the probability of another GW data ($d_{\rm GW2}$) on the basis of Equation~(\ref{eq:1})~\citep{2021ApJ...909..218A}:
\begin{equation}\begin{aligned}
    p&(H_0|d_{\rm GW1},d_{\rm GW2},d_{\rm EM}) \\
    \propto &
    \frac{p(H_0)}{\beta^2(H_0)}\,
    \int
    p(d_{\rm GW1}|{\hat D_L}(z_{\rm EM},H_0),\Omega_{\rm EM})\,\\&\times
    p(d_{\rm GW2}|{\hat D_L}(z_{\rm EM},H_0),\Omega_{\rm EM})\,
    p(D_L)\,p(\Omega)\,
    {\rm d}D_L\,{\rm d}\Omega \,.
    \label{eq:2}
\end{aligned}\end{equation}

\begin{figure}
\centering
\includegraphics[width=8cm]{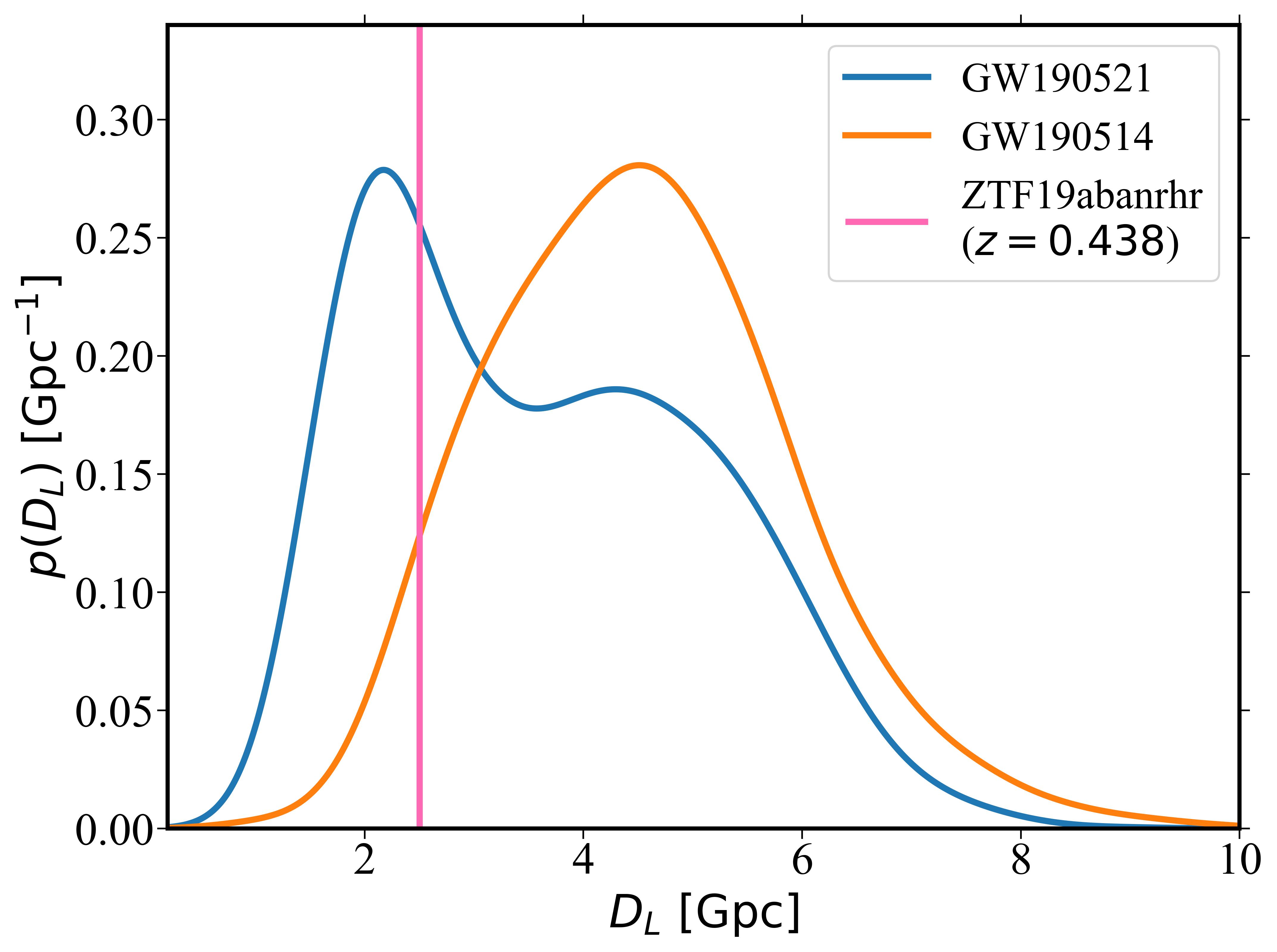}
\caption{
The probability density distribution of the luminosity distances for GW190514 (orange) and GW190521 (blue) along the line of sight to ZTF19abanrhr. The hotpink line represents the luminosity distance of ZTF19abanrhr is derived from the redshift
$z = 0.438$ of AGN J124942.3 + 344929, assuming the Planck18 cosmology~\citep{2020A&A...641A...6P}.
}
\label{fig1} 
\end{figure}

Here, we adopt two different prior choices for $H_0$: (1) a flat prior in the ranges $H_0=[10,180]\,\ksM$, and (2) GW170817 prior $H_0=70.0_{-8.0}^{+12.0}\,\ksM$ (the maximum a posteriori value and minimal 68.3\% credible interval) that is the inferred $H_0$ posterior from GW170817~\citep{2017Natur.551...85A}. We assume a flat $\Lambda$CDM and the matter density $\Omega_m=0.315$ from Planck18~\citep{2020A&A...641A...6P}.
We use gravitational wave datasets for GW190514 and GW190521 from the GWTC-2.1 release~\citep{2024PhRvD.109b2001A}, labeled as ``cosm'' and ``IMRPhenomXPHM''. 
We neglect uncertainties on the sky position (right ascension $RA. = 192.426257216$, declination $Dec.=34.824708900$) and redshift ($z = 0.438$) for the flare ZTF19abanrhr due to the better precision\footnote{\url{https://skyserver.sdss.org/dr12/en/tools/explore/Summary.aspx?id=1237665128546631763}} with respect to GW posterior distributions~(\citealp{2024PhRvD.109b2001A}, see Figure~\ref{fig1}). 
For instance, we show the probability distribution of the luminosity distances for GW190514 and GW190521 along the line of sight to ZTF19abanrhr in Figure~\ref{fig1}, with the luminosity distance of ZTF19abanrhr for comparison.

\subsection{Environmental effect analysis}\label{sec:EE}
Following~\citet{2023PhRvD.108l3039M}, the gravitational redshift is
\begin{equation}\label{grav}
z_{\rm grav} = 
\left(1-\frac{R_{\rm s}}{r}\right)^{1/2} - 1\,,
\end{equation}
where $R_{\rm s}$ represents the Schwarzschild radius of the supermassive black hole, and $r$ is the distance between the binary black hole and the supermassive black hole. Note that here assumes a non-spinning supermassive black hole because the effect of spin becomes negligible~\citep{2019ApJ...883L...7L,2019PhRvD..99j3005F,2022PhRvD.106j3040C}. The relativistic redshift is
\begin{equation}\label{rel}
z_{\rm rel} = 
\gamma \left[1+v\,{\rm cos}(\phi)\right] - 1 \,,
\end{equation}
where $\gamma = (1-v^2)^{-1/2}$ is the Lorentz factor, $v$ is the magnitude of the velocity, and $\phi$ is the viewing angle between the velocity and the line of sight in the observer frame. Assuming that the binary black hole is on a circular orbit around a non-spinning supermassive black hole~\citep{2020MNRAS.499.2608F}, its velocity is
\begin{equation}
v = 
\frac{1}{\sqrt{2\,(r/R_{\rm s}-1)}} \,.
\end{equation}

In this case, the effective distance ($D_L^{\rm eff}$) of the source is influenced by various redshift effects~\citep{2023PhRvD.107d3027T},
\begin{equation}\begin{aligned}
    D_L^{\rm eff} = 
    (1+z_{\rm grav})\,(1+z_{\rm rel})^2\,D_L,
    \label{eq:3}
\end{aligned}\end{equation}
where $D_L = (1+z_{\rm c})\,D_{\rm com}$ is the luminosity distance of the source with $D_{\rm com}$ the comoving distance between the source and the observer.
Thus, the environmental effects can be accounted for by simply replacing $D_L$ with $D_L^{\rm eff}$~\citep{2023PhRvD.108l3039M}. 

We adopt that the viewing angle $\phi$ follows an uniform distribution in ${\rm cos}(\phi)$. We assume that the distance $r$ is uniformly distributed between 24.5 and 331 Schwarzschild radii, where ``migration traps'' in AGN disks are expected to exist~\citep{2016ApJ...819L..17B}.

\section{Results and Discussion}\label{sec:RD}

\begin{figure}
\centering
\includegraphics[width=8cm]{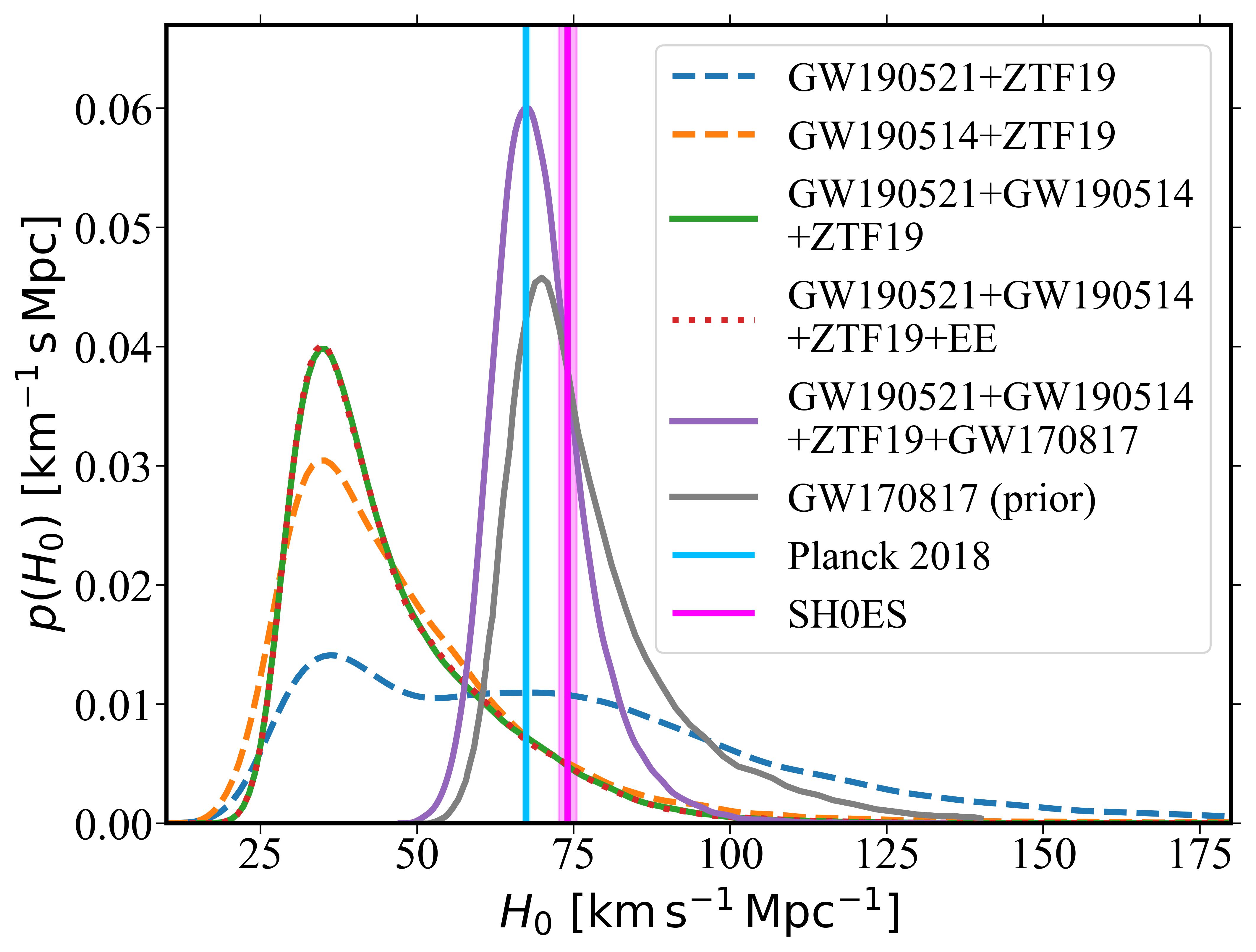}
\caption{
The posterior probability density distribution of $H_0$.
The green solid (red dotted) curve uses the $H_0$ distribution from the hierarchical triple merger GW190514-GW190521 with its EM counterpart ZTF19abanrhr without (with) environmental effects, as a prior assumes a flat prior on $H_0$. 
The purple solid curve represents the $H_0$ distribution for GW190514-GW190521 with ZTF19abanrhr, assuming the GW170817 prior (gray solid) for $H_0$.
As comparison, the blue (orange) dashed curve shows the $H_0$ distribution from GW190521 (GW190514) with ZTF19abanrhr. 
The $H_0$ estimates from both cosmic microwave background by Planck~(\citealp{2020A&A...641A...6P}, deepskyblue) and Type Ia supernova by SH0ES~(\citealp{2019ApJ...876...85R}, magenta) are presented for comparison.
}
\label{fig2} 
\end{figure}

Figure~\ref{fig2} displays the posterior probability density distributions of $H_0$ from various sources and assumptions. 
We find that our inferred $H_0$ distribution is $H_0=40.9_{-8.9}^{+19.3}\,\ksM$ (median and 68\% credible interval) from hierarchical triple merger GW190514-GW190521 with a flat prior on $H_0$. 
Adopting the GW170817 prior yields $H_0=68.8_{-6.0}^{+7.7}\,\ksM$. This combination improves the measurement of $H_0$. In particular, the combined approach reduces the relative uncertainty ($\sigma_{H_0}/H_0$) from $\sim$14.3\% (GW170817 alone,~\citealp{2017Natur.551...85A}) to $\sim$10.0\%. 
We see that the inferred $H_0$ values are consistent with the estimates from Planck18 ($H_0=67.4\pm0.5\,\ksM$,~\citealp{2020A&A...641A...6P}) and SH0ES ($H_0=74.03\pm1.42\,\ksM$,~\citealp{2019ApJ...876...85R}), while exhibiting preferred alignment with Planck18.

When incorporating environmental effects, the hierarchical triple merger analysis maintains nearly identical values $H_0=40.9_{-8.9}^{+19.2}\,\ksM$ compared to the baseline measurement. The relative uncertainty remains stable at $\sim$34.4-34.5\%, indicating minimal impact ($\lesssim$0.1\%) from environmental factors under current observational precision. This marginal influence may stem from the (typically) inherent $\pm$30\% luminosity distance uncertainty in current binary black hole merger detections~(although~\citealp{2025ApJ...984...63L}). Notably, extreme scenarios involving mergers within a few Schwarzschild radii~\citep{2021MNRAS.505.1324P} or orbital alignments with the line of sight may require revised environmental considerations.

Individual event analyses yield $H_0=66.7_{-29.8}^{+38.8}\,\ksM$ for GW190521 and $42.7_{-11.6}^{+20.8}\,\ksM$ for GW190514 under flat priors on $H_0$. To our knowledge, this constitutes the first measurement using GW190514 (or GW190514-GW190521).
The combined GW190514-GW190521 analysis achieves $\sigma_{H_0}/H_0\sim34.5\%$, demonstrating measurable improvement over individual uncertainties with $\sim$51.4\% for GW190521 and $\sim$37.9\% for GW190514. This underscores the value of hierarchical triple merger analysis with its potential EM counterpart

We note that, while the association between the GW190514-GW190521 merger pair and the optical flare ZTF19abanrhr is statistically plausible based on spatial and mass coincidence~\citep {PhysRevD.111.103016}, further validation of this association (GW190514-GW190521-ZTF19abanrhr) is needed. For example, this can be achieved through constraints imposed by factors such as the time delay between the two GW signals and the eccentricity of the source~\citep {2017ApJ...835..165B,2018ApJ...866...66M,2019ApJ...877...87Z,2020ApJ...903L...5R,2022NatAs...6..344G,2022Natur.603..237S,2022ApJ...928L...1L,2024arXiv240101743H,2025arXiv250110703W}.
Our analysis adopts this scenario as a working hypothesis to explore environmental effects near supermassive black holes on $H_0$. If future studies disprove this association, our results on redshift corrections would still hold as a general framework for GW mergers in AGNs, though the specific $H_0$ value derived here would require revision.

The estimated value of $H_0$ is $\sim$$40\,\rm km\,s^{-1}\,Mpc^{-1}$ when it comes to GW190514, which seems to be far from the well-known value~\citep{2019ApJ...876...85R,2020A&A...641A...6P}.
In addition to the possible false multimessenger, the main reason may be the large uncertainty in luminosity distance measured for GW sources. At the current GW detector sensitivity, this deviated $H_0$ value is normal and is also common in previous literature, as analyzed below.
Our results align with previous GW190521 studies employing different methodologies. Previous studies report GW190521-derived $H_0$ values spanning $\sim$43-102$\,\ksM$ under different physical assumptions: 
lower-range measurements ($H_0=48_{-10}^{+23}\,\ksM$,~\citealp{2022MNRAS.513.2152C} to $50.4_{-19.5}^{+28.1}\,\ksM$,~\citealp{2020arXiv200914199M}) correspond to standard waveform (NRSur7dq4) analyses from ~\citet{2020PhRvL.125j1102A}, higher values ($H_0=73_{-15}^{+24}\,\ksM$,~\citealp{2021arXiv211212481C} to $H_0=102_{-25}^{+27}\,\ksM$,~\citealp{2023PhRvD.108l3039M}) arise from mass ratio priors or data selection differences, while $H_0=68.8_{-25.5}^{+45.7}\,\ksM$~\citep{2021ApJ...908L..34G} reflects eccentric orbital assumptions.

We note that for GW events occurring at distances exceeding tens of Schwarzschild radii from the supermassive black hole, the gravitational and relativistic redshifts are intuitively negligible, being significantly smaller than the uncertainty in luminosity distance estimation from GW observation.
Despite the minimal magnitude of these environmental effects, verifying their insignificance is crucial to ensure the accuracy of $H_0$ measurements. In particular, we suggest that these effects might only become appreciable and non-ignorable in extreme scenarios, such as when events are located within a few Schwarzschild radii.
In addition, although the current uncertainty in luminosity distance measurements for GW events is substantial, future detectors are expected to enhance measurement precision. Consequently, these seemingly minor effects could gain critical importance in the context of more accurate observations.
Demonstrating the insignificance of redshift corrections in this case supports the broader use of AGN-associated mergers as `clean' laboratories for cosmology, free from significant environmental contamination.

\section{Conclusions}\label{sec:conslusions}

We investigate environmental effects' impact on Hubble constant ($H_0$) measurements for stellar-mass binary black holes residing in AGN environments near the supermassive black hole. We use the potential hierarchical triple merger GW190514-GW190521 detected by LIGO-Virgo with its associated EM counterpart ZTF19abanrhr detected by ZTF as a multimessenger case study~\citep{PhysRevD.111.103016}, and therefore quantify these influences. Our findings reveal:
\begin{enumerate}
    \item Environmental effects induce negligible $H_0$ corrections ($\sim$0.1\%) for binaries in AGNs , suggesting their minimal impact under current observational precision (even for future detectors, see~\citealp{2025PhRvL.134h1402S,2025arXiv250324084Z}).
    \item Combined analysis of GW190514-GW190521 and ZTF19abanrhr yields $H_0=40.9_{-8.9}^{+19.3}\,\ksM$ (flat prior) , improving to $H_0=68.8_{-6.0}^{+7.7}\,\ksM$ when incorporating the GW170817 prior.
    \item Individual event measurements show $H_0=66.7_{-29.8}^{+38.8}\,\ksM$ for GW190521 and $42.7_{-11.6}^{+20.8}\,\ksM$ for GW190514, demonstrating the precision enhancement from the hierarchical triple merger GW190514-GW190521.
\end{enumerate}

The observed improvement~\citep{2010CQGra..27s4002P,2010CQGra..27h4007P,2017CQGra..34d4001A,2020LRR....23....3A} stems from AGN environments' multimessenger potential~\citep{2020PhRvL.124y1102G,2023ApJ...942...99G,2022MNRAS.514.2092V,2023MNRAS.526.6031V,2024arXiv240505318V,2024arXiv240721568V,2024PhRvD.110l3029C}: gas-rich disks surrounding merging black holes may produce detectable EM flares~(e.g.,~\citealp{2019ApJ...884L..50M,2021ApJ...916L..17W,2023ApJ...955...23T,2023MNRAS.524.6015L,2024ApJ...961..206C,2024arXiv240610904Z}). With upgraded (or next-generation) GW detectors~\citep{2010CQGra..27s4002P,2010CQGra..27h4007P,2017CQGra..34d4001A,2020LRR....23....3A} expected to identify more AGN-channel mergers accompanied by EM counterparts, we anticipate significantly refined $H_0$ constraints through multimessenger synergies~\citep{2024MNRAS.531.3679A,2024PhRvD.110h3005B}.

\section{Acknowledgments}
We would like to thank the referees and Ilya Mandel for their valuable comments and suggestions, which considerably improved the original manuscript.
This work is supported by National Key R$\&$D Program of China (2020YFC2201400). 
This research has made use of data or software obtained from the Gravitational Wave Open Science Center (\url{https://gwosc.org}), a service of the LIGO Scientific Collaboration, the Virgo Collaboration, and KAGRA. This material is based upon work supported by NSF's LIGO Laboratory which is a major facility fully funded by the National Science Foundation, as well as the Science and Technology Facilities Council (STFC) of the United Kingdom, the Max-Planck-Society (MPS), and the State of Niedersachsen/Germany for support of the construction of Advanced LIGO and construction and operation of the GEO600 detector. Additional support for Advanced LIGO was provided by the Australian Research Council. Virgo is funded, through the European Gravitational Observatory (EGO), by the French Centre National de Recherche Scientifique (CNRS), the Italian Istituto Nazionale di Fisica Nucleare (INFN) and the Dutch Nikhef, with contributions by institutions from Belgium, Germany, Greece, Hungary, Ireland, Japan, Monaco, Poland, Portugal, Spain. KAGRA is supported by Ministry of Education, Culture, Sports, Science and Technology (MEXT), Japan Society for the Promotion of Science (JSPS) in Japan; National Research Foundation (NRF) and Ministry of Science and ICT (MSIT) in Korea; Academia Sinica (AS) and National Science and Technology Council (NSTC) in Taiwan.
This analysis was made possible following software packages:
NumPy~\citep{harris2020array}, 
SciPy~\citep{2020SciPy-NMeth}, 
Matplotlib~\citep{2007CSE.....9...90H}, 
emcee~\citep{2013PASP..125..306F}, 
IPython~\citep{2007CSE.....9c..21P},
corner~\citep{2016JOSS....1...24F},
seaborn~\citep{Waskom2021},
and Astropy~\citep{2022ApJ...935..167A}.


\begin{thebibliography}{}
\expandafter\ifx\csname natexlab\endcsname\relax\def\natexlab#1{#1}\fi
\providecommand{\url}[1]{\href{#1}{#1}}
\providecommand{\dodoi}[1]{doi:~\href{http://doi.org/#1}{\nolinkurl{#1}}}
\providecommand{\doeprint}[1]{\href{http://ascl.net/#1}{\nolinkurl{http://ascl.net/#1}}}
\providecommand{\doarXiv}[1]{\href{https://arxiv.org/abs/#1}{\nolinkurl{https://arxiv.org/abs/#1}}}

\bibitem[{B.~P. {Abbott} {et~al.}(2017{\natexlab{a}}){Abbott}, {Abbott}, {Abbott}, {Acernese}, {Ackley}, {Adams}, {Adams}, {Addesso}, {Adhikari}, {Adya}, {Affeldt}, {Afrough}, {Agarwal}, {Agathos}, {Agatsuma}, {Aggarwal}, {Aguiar}, {Aiello}, {Ain}, {Ajith}, {Allen}, {Allen}, {Allocca}, {Altin}, {Amato}, {Ananyeva}, {Anderson}, {Anderson}, {Angelova}, {Antier}, {Appert}, {Arai}, {Araya}, {Areeda}, {Arnaud}, {Arun}, {Ascenzi}, {Ashton}, {Ast}, {Aston}, {Astone}, {Atallah}, {Aufmuth}, {Aulbert}, {Aultoneal}, {Austin}, {Avila-Alvarez}, {Babak}, {Bacon}, {Bader}, {Bae}, {Baker}, {Baldaccini}, {Ballardin}, {Ballmer}, {Banagiri}, {Barayoga}, {Barclay}, {Barish}, {Barker}, {Barkett}, {Barone}, {Barr}, {Barsotti}, {Barsuglia}, {Barta}, {Bartlett}, {Bartos}, {Bassiri}, {Basti}, {Batch}, {Bawaj}, {Bayley}, {Bazzan}, {B{\'e}csy}, {Beer}, {Bejger}, {Belahcene}, {Bell}, {Berger}, {Bergmann}, {Bero}, {Berry}, {Bersanetti}, {Bertolini}, {Betzwieser}, {Bhagwat}, {Bhandare}, {Bilenko}, {Billingsley}, {Billman}, {Birch},
  {Birney}, {Birnholtz}, {Biscans}, {Biscoveanu}, {Bisht}, {Bitossi}, {Biwer}, {Bizouard}, {Blackburn}, {Blackman}, {Blair}, {Blair}, {Blair}, {Bloemen}, {Bock}, {Bode}, {Boer}, {Bogaert}, {Bohe}, {Bondu}, {Bonilla}, {Bonnand}, {Boom}, {Bork}, {Boschi}, {Bose}, {Bossie}, {Bouffanais}, {Bozzi}, {Bradaschia}, {Brady}, {Branchesi}, {Brau}, {Briant}, {Brillet}, {Brinkmann}, {Brisson}, {Brockill}, {Broida}, {Brooks}, {Brown}, {Brown}, {Brunett}, {Buchanan}, {Buikema}, {Bulik}, {Bulten}, {Buonanno}, {Buskulic}, {Buy}, {Byer}, {Cabero}, {Cadonati}, {Cagnoli}, {Cahillane}, {Bustillo}, {Callister}, {Calloni}, {Camp}, {Canepa}, {Canizares}, {Cannon}, {Cao}, {Cao}, {Capano}, {Capocasa}, {Carbognani}, {Caride}, {Carney}, {Diaz}, {Casentini}, {Caudill}, {Cavagli{\`a}}, {Cavalier}, {Cavalieri}, {Cella}, {Cepeda}, {Cerd{\'a}-Dur{\'a}n}, {Cerretani}, {Cesarini}, {Chamberlin}, {Chan}, {Chao}, {Charlton}, {Chase}, {Chassande-Mottin}, {Chatterjee}, {Chatziioannou}, {Cheeseboro}, {Chen}, {Chen}, {Chen}, {Cheng}, {Chia},
  {Chincarini}, {Chiummo}, {Chmiel}, {Cho}, {Cho}, {Chow}, {Christensen}, {Chu}, {Chua}, {Chua}, {Chung}, {Chung}, {Ciani}, \& {Ciolfi}}]{2017Natur.551...85A}
{Abbott}, B.~P., {Abbott}, R., {Abbott}, T.~D., {et~al.} 2017{\natexlab{a}}, \bibinfo{title}{{A gravitational-wave standard siren measurement of the Hubble constant},} \nat, 551, 85, \dodoi{10.1038/nature24471}

\bibitem[{B.~P. {Abbott} {et~al.}(2017{\natexlab{b}}){Abbott}, {Abbott}, {Abbott}, {Acernese}, {Ackley}, {Adams}, {Adams}, {Addesso}, {Adhikari}, {Adya}, {Affeldt}, {Afrough}, {Agarwal}, {Agathos}, {Agatsuma}, {Aggarwal}, {Aguiar}, {Aiello}, {Ain}, {Ajith}, {Allen}, {Allen}, {Allocca}, {Altin}, {Amato}, {Ananyeva}, {Anderson}, {Anderson}, {Angelova}, {Antier}, {Appert}, {Arai}, {Araya}, {Areeda}, {Arnaud}, {Arun}, {Ascenzi}, {Ashton}, {Ast}, {Aston}, {Astone}, {Atallah}, {Aufmuth}, {Aulbert}, {AultONeal}, {Austin}, {Avila-Alvarez}, {Babak}, {Bacon}, {Bader}, {Bae}, {Bailes}, {Baker}, {Baldaccini}, {Ballardin}, {Ballmer}, {Banagiri}, {Barayoga}, {Barclay}, {Barish}, {Barker}, {Barkett}, {Barone}, {Barr}, {Barsotti}, {Barsuglia}, {Barta}, {Barthelmy}, {Bartlett}, {Bartos}, {Bassiri}, {Basti}, {Batch}, {Bawaj}, {Bayley}, {Bazzan}, {B{\'e}csy}, {Beer}, {Bejger}, {Belahcene}, {Bell}, {Berger}, {Bergmann}, {Bernuzzi}, {Bero}, {Berry}, {Bersanetti}, {Bertolini}, {Betzwieser}, {Bhagwat}, {Bhandare}, {Bilenko},
  {Billingsley}, {Billman}, {Birch}, {Birney}, {Birnholtz}, {Biscans}, {Biscoveanu}, {Bisht}, {Bitossi}, {Biwer}, {Bizouard}, {Blackburn}, {Blackman}, {Blair}, {Blair}, {Blair}, {Bloemen}, {Bock}, {Bode}, {Boer}, {Bogaert}, {Bohe}, {Bondu}, {Bonilla}, {Bonnand}, {Boom}, {Bork}, {Boschi}, {Bose}, {Bossie}, {Bouffanais}, {Bozzi}, {Bradaschia}, {Brady}, {Branchesi}, {Brau}, {Briant}, {Brillet}, {Brinkmann}, {Brisson}, {Brockill}, {Broida}, {Brooks}, {Brown}, {Brown}, {Brunett}, {Buchanan}, {Buikema}, {Bulik}, {Bulten}, {Buonanno}, {Buskulic}, {Buy}, {Byer}, {Cabero}, {Cadonati}, {Cagnoli}, {Cahillane}, {Calder{\'o}n Bustillo}, {Callister}, {Calloni}, {Camp}, {Canepa}, {Canizares}, {Cannon}, {Cao}, {Cao}, {Capano}, {Capocasa}, {Carbognani}, {Caride}, {Carney}, {Carullo}, {Casanueva Diaz}, {Casentini}, {Caudill}, {Cavagli{\`a}}, {Cavalier}, {Cavalieri}, {Cella}, {Cepeda}, {Cerd{\'a}-Dur{\'a}n}, {Cerretani}, {Cesarini}, {Chamberlin}, {Chan}, {Chao}, {Charlton}, {Chase}, {Chassande-Mottin}, {Chatterjee},
  {Chatziioannou}, {Cheeseboro}, {Chen}, {Chen}, {Chen}, {Cheng}, {Chia}, {Chincarini}, {Chiummo}, {Chmiel}, {Cho}, {Cho}, {Chow}, {Christensen}, {Chu}, {Chua}, \& {Chua}}]{2017PhRvL.119p1101A}
{Abbott}, B.~P., {Abbott}, R., {Abbott}, T.~D., {et~al.} 2017{\natexlab{b}}, \bibinfo{title}{{GW170817: Observation of Gravitational Waves from a Binary Neutron Star Inspiral},} \prl, 119, 161101, \dodoi{10.1103/PhysRevLett.119.161101}

\bibitem[{B.~P. {Abbott} {et~al.}(2017{\natexlab{c}}){Abbott}, {Abbott}, {Abbott}, {Abernathy}, {Ackley}, {Adams}, {Addesso}, {Adhikari}, {Adya}, {Affeldt}, \& et~al.}]{2017CQGra..34d4001A}
{Abbott}, B.~P., {Abbott}, R., {Abbott}, T.~D., {et~al.} 2017{\natexlab{c}}, \bibinfo{title}{{Exploring the sensitivity of next generation gravitational wave detectors},} Classical and Quantum Gravity, 34, 044001, \dodoi{10.1088/1361-6382/aa51f4}

\bibitem[{B.~P. {Abbott} {et~al.}(2020){Abbott}, {Abbott}, {Abbott}, {Abraham}, {Acernese}, {Ackley}, {Adams}, {Adya}, {Affeldt}, {Agathos}, {Agatsuma}, {Aggarwal}, {Aguiar}, {Aiello}, {Ain}, {Ajith}, {Akutsu}, {Allen}, {Allocca}, {Aloy}, {Altin}, {Amato}, {Ananyeva}, {Anderson}, {Anderson}, {Ando}, {Angelova}, {Antier}, {Appert}, {Arai}, {Arai}, {Arai}, {Araki}, {Araya}, {Araya}, {Areeda}, {Ar{\`e}ne}, {Aritomi}, {Arnaud}, {Arun}, {Ascenzi}, {Ashton}, {Aso}, {Aston}, {Astone}, {Aubin}, {Aufmuth}, {Aultoneal}, {Austin}, {Avendano}, {Avila-Alvarez}, {Babak}, {Bacon}, {Badaracco}, {Bader}, {Bae}, {Bae}, {Baiotti}, {Bajpai}, {Baker}, {Baldaccini}, {Ballardin}, {Ballmer}, {Banagiri}, {Barayoga}, {Barclay}, {Barish}, {Barker}, {Barkett}, {Barnum}, {Barone}, {Barr}, {Barsotti}, {Barsuglia}, {Barta}, {Bartlett}, {Barton}, {Bartos}, {Bassiri}, {Basti}, {Bawaj}, {Bayley}, {Bazzan}, {B{\'e}csy}, {Bejger}, {Belahcene}, {Bell}, {Beniwal}, {Berger}, {Bergmann}, {Bernuzzi}, {Bero}, {Berry}, {Bersanetti}, {Bertolini},
  {Betzwieser}, {Bhandare}, {Bidler}, {Bilenko}, {Bilgili}, {Billingsley}, {Birch}, {Birney}, {Birnholtz}, {Biscans}, {Biscoveanu}, {Bisht}, {Bitossi}, {Bizouard}, {Blackburn}, {Blair}, {Blair}, {Blair}, {Bloemen}, {Bode}, {Boer}, {Boetzel}, {Bogaert}, {Bondu}, {Bonilla}, {Bonnand}, {Booker}, {Boom}, {Booth}, {Bork}, {Boschi}, {Bose}, {Bossie}, {Bossilkov}, {Bosveld}, {Bouffanais}, {Bozzi}, {Bradaschia}, {Brady}, {Bramley}, {Branchesi}, {Brau}, {Briant}, {Briggs}, {Brighenti}, {Brillet}, {Brinkmann}, {Brisson}, {Brockill}, {Brooks}, {Brown}, {Brown}, {Brunett}, {Buikema}, {Bulik}, {Bulten}, {Buonanno}, {Buskulic}, {Buy}, {Byer}, {Cabero}, {Cadonati}, {Cagnoli}, {Cahillane}, {Bustillo}, {Callister}, {Calloni}, {Camp}, {Campbell}, {Canepa}, {Cannon}, {Cannon}, {Cao}, {Cao}, {Capocasa}, {Carbognani}, {Caride}, {Carney}, {Carullo}, {Diaz}, {Casentini}, {Caudill}, {Cavagli{\`a}}, {Cavalier}, {Cavalieri}, {Cella}, {Cerd{\'a}-Dur{\'a}n}, {Cerretani}, {Cesarini}, {Chaibi}, {Chakravarti}, {Chamberlin}, {Chan}, {Chan},
  {Chao}, {Charlton}, {Chase}, {Chassande-Mottin}, {Chatterjee}, {Chaturvedi}, {Chatziioannou}, {Cheeseboro}, {Chen}, {Chen}, \& {Chen}}]{2020LRR....23....3A}
{Abbott}, B.~P., {Abbott}, R., {Abbott}, T.~D., {et~al.} 2020, \bibinfo{title}{{Prospects for observing and localizing gravitational-wave transients with Advanced LIGO, Advanced Virgo and KAGRA},} Living Reviews in Relativity, 23, 3, \dodoi{10.1007/s41114-020-00026-9}

\bibitem[{B.~P. {Abbott} {et~al.}(2021){Abbott}, {Abbott}, {Abbott}, {Abraham}, {Acernese}, {Ackley}, {Adams}, {Adhikari}, {Adya}, {Affeldt}, {Agathos}, {Agatsuma}, {Aggarwal}, {Aguiar}, {Aiello}, {Ain}, {Ajith}, {Allen}, {Allocca}, {Aloy}, {Altin}, {Amato}, {Anand}, {Ananyeva}, {Anderson}, {Anderson}, {Angelova}, {Antier}, {Appert}, {Arai}, {Araya}, {Areeda}, {Ar{\`e}ne}, {Arnaud}, {Aronson}, {Arun}, {Ascenzi}, {Ashton}, {Aston}, {Astone}, {Aubin}, {Aufmuth}, {AultONeal}, {Austin}, {Avendano}, {Avila-Alvarez}, {Babak}, {Bacon}, {Badaracco}, {Bader}, {Bae}, {Baird}, {Baker}, {Baldaccini}, {Ballardin}, {Ballmer}, {Bals}, {Banagiri}, {Barayoga}, {Barbieri}, {Barclay}, {Barish}, {Barker}, {Barkett}, {Barnum}, {Barone}, {Barr}, {Barsotti}, {Barsuglia}, {Barta}, {Bartlett}, {Bartos}, {Bassiri}, {Basti}, {Bawaj}, {Bayley}, {Bazzan}, {B{\'e}csy}, {Bejger}, {Belahcene}, {Bell}, {Beniwal}, {Benjamin}, {Berger}, {Bergmann}, {Bernuzzi}, {Berry}, {Bersanetti}, {Bertolini}, {Betzwieser}, {Bhandare}, {Bidler}, {Biggs},
  {Bilenko}, {Bilgili}, {Billingsley}, {Birney}, {Birnholtz}, {Biscans}, {Bischi}, {Biscoveanu}, {Bisht}, {Bitossi}, {Bizouard}, {Blackburn}, {Blackman}, {Blair}, {Blair}, {Blair}, {Bloemen}, {Bobba}, {Bode}, {Boer}, {Boetzel}, {Bogaert}, {Bondu}, {Bonnand}, {Booker}, {Boom}, {Bork}, {Boschi}, {Bose}, {Bossilkov}, {Bosveld}, {Bouffanais}, {Bozzi}, {Bradaschia}, {Brady}, {Bramley}, {Branchesi}, {Brau}, {Breschi}, {Briant}, {Briggs}, {Brighenti}, {Brillet}, {Brinkmann}, {Brockill}, {Brooks}, {Brooks}, {Brown}, {Brunett}, {Buikema}, {Bulik}, {Bulten}, {Buonanno}, {Buskulic}, {Buy}, {Byer}, {Cabero}, {Cadonati}, {Cagnoli}, {Cahillane}, {Calder{\'o}n Bustillo}, {Callister}, {Calloni}, {Camp}, {Campbell}, {Canepa}, {Cannon}, {Cao}, {Cao}, {Carapella}, {Carbognani}, {Caride}, {Carney}, {Carullo}, {Casanueva Diaz}, {Casentini}, {Caudill}, {Cavagli{\`a}}, {Cavalier}, {Cavalieri}, {Cella}, {Cerd{\'a}-Dur{\'a}n}, {Cesarini}, {Chaibi}, {Chakravarti}, {Chamberlin}, {Chan}, {Chao}, {Charlton}, {Chase}, {Chassande-Mottin},
  {Chatterjee}, {Chaturvedi}, {Cheeseboro}, {Chen}, {Chen}, {Chen}, {Cheng}, {Cheong}, {Chia}, {Chiadini}, {Chincarini}, {Chiummo}, {Cho}, {Cho}, {Cho}, \& {Christensen}}]{2021ApJ...909..218A}
{Abbott}, B.~P., {Abbott}, R., {Abbott}, T.~D., {et~al.} 2021, \bibinfo{title}{{A Gravitational-wave Measurement of the Hubble Constant Following the Second Observing Run of Advanced LIGO and Virgo},} \apj, 909, 218, \dodoi{10.3847/1538-4357/abdcb7}

\bibitem[{R. {Abbott} {et~al.}(2020{\natexlab{a}}){Abbott}, {Abbott}, {Abraham}, {Acernese}, {Ackley}, {Adams}, {Adhikari}, {Adya}, {Affeldt}, {Agathos}, \& et~al.}]{2020PhRvL.125j1102A}
{Abbott}, R., {Abbott}, T.~D., {Abraham}, S., {et~al.} 2020{\natexlab{a}}, \bibinfo{title}{{GW190521: A Binary Black Hole Merger with a Total Mass of 150 M$_{{\ensuremath{\odot}}}$},} \prl, 125, 101102, \dodoi{10.1103/PhysRevLett.125.101102}

\bibitem[{R. {Abbott} {et~al.}(2020{\natexlab{b}}){Abbott}, {Abbott}, {Abraham}, {Acernese}, {Ackley}, {Adams}, {Adhikari}, {Adya}, {Affeldt}, {Agathos}, \& et~al.}]{2020ApJ...900L..13A}
{Abbott}, R., {Abbott}, T.~D., {Abraham}, S., {et~al.} 2020{\natexlab{b}}, \bibinfo{title}{{Properties and Astrophysical Implications of the 150 M$_{{\ensuremath{\odot}}}$ Binary Black Hole Merger GW190521},} \apjl, 900, L13, \dodoi{10.3847/2041-8213/aba493}

\bibitem[{R. {Abbott} {et~al.}(2021){Abbott}, {Abbott}, {Abraham}, {Acernese}, {Ackley}, {Adams}, {Adams}, {Adhikari}, {Adya}, {Affeldt}, \& et~al.}]{2021PhRvX..11b1053A}
{Abbott}, R., {Abbott}, T.~D., {Abraham}, S., {et~al.} 2021, \bibinfo{title}{{GWTC-2: Compact Binary Coalescences Observed by LIGO and Virgo during the First Half of the Third Observing Run},} Physical Review X, 11, 021053, \dodoi{10.1103/PhysRevX.11.021053}

\bibitem[{R. {Abbott} {et~al.}(2024){Abbott}, {Abbott}, {Acernese}, {Ackley}, {Adams}, {Adhikari}, {Adhikari}, {Adya}, {Affeldt}, {Agarwal}, \& et~al.}]{2024PhRvD.109b2001A}
{Abbott}, R., {Abbott}, T.~D., {Acernese}, F., {et~al.} 2024, \bibinfo{title}{{GWTC-2.1: Deep extended catalog of compact binary coalescences observed by LIGO and Virgo during the first half of the third observing run},} \prd, 109, 022001, \dodoi{10.1103/PhysRevD.109.022001}

\bibitem[{F. {Acernese} {et~al.}(2015){Acernese}, {Agathos}, {Agatsuma}, {Aisa}, {Allemandou}, {Allocca}, {Amarni}, {Astone}, {Balestri}, {Ballardin}, \& et~al.}]{2015CQGra..32b4001A}
{Acernese}, F., {Agathos}, M., {Agatsuma}, K., {et~al.} 2015, \bibinfo{title}{{Advanced Virgo: a second-generation interferometric gravitational wave detector},} Classical and Quantum Gravity, 32, 024001, \dodoi{10.1088/0264-9381/32/2/024001}

\bibitem[{L.~M.~B. {Alves} {et~al.}(2024){Alves}, {Sullivan}, {Yang}, {Gayathri}, {M{\'a}rka}, {M{\'a}rka}, \& {Bartos}}]{2024MNRAS.531.3679A}
{Alves}, L. M.~B., {Sullivan}, A.~G., {Yang}, Y., {et~al.} 2024, \bibinfo{title}{{Determining the Hubble constant with AGN-assisted black hole mergers},} \mnras, 531, 3679, \dodoi{10.1093/mnras/stae1360}

\bibitem[{Y. {Aso} {et~al.}(2013){Aso}, {Michimura}, {Somiya}, {Ando}, {Miyakawa}, {Sekiguchi}, {Tatsumi}, \& {Yamamoto}}]{2013PhRvD..88d3007A}
{Aso}, Y., {Michimura}, Y., {Somiya}, K., {et~al.} 2013, \bibinfo{title}{{Interferometer design of the KAGRA gravitational wave detector},} \prd, 88, 043007, \dodoi{10.1103/PhysRevD.88.043007}

\bibitem[{ {Astropy Collaboration} {et~al.}(2022){Astropy Collaboration}, {Price-Whelan}, {Lim}, {Earl}, {Starkman}, {Bradley}, {Shupe}, {Patil}, {Corrales}, {Brasseur}, {N{\"o}the}, {Donath}, {Tollerud}, {Morris}, {Ginsburg}, {Vaher}, {Weaver}, {Tocknell}, {Jamieson}, {van Kerkwijk}, {Robitaille}, {Merry}, {Bachetti}, {G{\"u}nther}, {Aldcroft}, {Alvarado-Montes}, {Archibald}, {B{\'o}di}, {Bapat}, {Barentsen}, {Baz{\'a}n}, {Biswas}, {Boquien}, {Burke}, {Cara}, {Cara}, {Conroy}, {Conseil}, {Craig}, {Cross}, {Cruz}, {D'Eugenio}, {Dencheva}, {Devillepoix}, {Dietrich}, {Eigenbrot}, {Erben}, {Ferreira}, {Foreman-Mackey}, {Fox}, {Freij}, {Garg}, {Geda}, {Glattly}, {Gondhalekar}, {Gordon}, {Grant}, {Greenfield}, {Groener}, {Guest}, {Gurovich}, {Handberg}, {Hart}, {Hatfield-Dodds}, {Homeier}, {Hosseinzadeh}, {Jenness}, {Jones}, {Joseph}, {Kalmbach}, {Karamehmetoglu}, {Ka{\l}uszy{\'n}ski}, {Kelley}, {Kern}, {Kerzendorf}, {Koch}, {Kulumani}, {Lee}, {Ly}, {Ma}, {MacBride}, {Maljaars}, {Muna}, {Murphy}, {Norman},
  {O'Steen}, {Oman}, {Pacifici}, {Pascual}, {Pascual-Granado}, {Patil}, {Perren}, {Pickering}, {Rastogi}, {Roulston}, {Ryan}, {Rykoff}, {Sabater}, {Sakurikar}, {Salgado}, {Sanghi}, {Saunders}, {Savchenko}, {Schwardt}, {Seifert-Eckert}, {Shih}, {Jain}, {Shukla}, {Sick}, {Simpson}, {Singanamalla}, {Singer}, {Singhal}, {Sinha}, {Sip{\H{o}}cz}, {Spitler}, {Stansby}, {Streicher}, {{\v{S}}umak}, {Swinbank}, {Taranu}, {Tewary}, {Tremblay}, {de Val-Borro}, {Van Kooten}, {Vasovi{\'c}}, {Verma}, {de Miranda Cardoso}, {Williams}, {Wilson}, {Winkel}, {Wood-Vasey}, {Xue}, {Yoachim}, {Zhang}, {Zonca}, \& {Astropy Project Contributors}}]{2022ApJ...935..167A}
{Astropy Collaboration}, {Price-Whelan}, A.~M., {Lim}, P.~L., {et~al.} 2022, \bibinfo{title}{{The Astropy Project: Sustaining and Growing a Community-oriented Open-source Project and the Latest Major Release (v5.0) of the Core Package},} \apj, 935, 167, \dodoi{10.3847/1538-4357/ac7c74}

\bibitem[{I. {Bartos} {et~al.}(2017){Bartos}, {Kocsis}, {Haiman}, \& {M{\'a}rka}}]{2017ApJ...835..165B}
{Bartos}, I., {Kocsis}, B., {Haiman}, Z., \& {M{\'a}rka}, S. 2017, \bibinfo{title}{{Rapid and Bright Stellar-mass Binary Black Hole Mergers in Active Galactic Nuclei},} \apj, 835, 165, \dodoi{10.3847/1538-4357/835/2/165}

\bibitem[{J.~M. {Bellovary} {et~al.}(2016){Bellovary}, {Mac Low}, {McKernan}, \& {Ford}}]{2016ApJ...819L..17B}
{Bellovary}, J.~M., {Mac Low}, M.-M., {McKernan}, B., \& {Ford}, K.~E.~S. 2016, \bibinfo{title}{{Migration Traps in Disks around Supermassive Black Holes},} \apjl, 819, L17, \dodoi{10.3847/2041-8205/819/2/L17}

\bibitem[{C.~R. {Bom} \& A. {Palmese}(2024){Bom} \& {Palmese}}]{2024PhRvD.110h3005B}
{Bom}, C.~R., \& {Palmese}, A. 2024, \bibinfo{title}{{Standard siren cosmology with gravitational waves from binary black hole mergers in active galactic nuclei},} \prd, 110, 083005, \dodoi{10.1103/PhysRevD.110.083005}

\bibitem[{T. {Cabrera} {et~al.}(2024){Cabrera}, {Palmese}, {Hu}, {O'Connor}, {Ford}, {McKernan}, {Andreoni}, {Ahumada}, {Amsellem}, {Busmann}, {Clark}, {Coughlin}, {Dadiani}, {Diaz}, {Graham}, {Gruen}, {Kunnumkai}, {Postiglione}, {Riffeser}, {Sommer}, \& {Valdes}}]{2024PhRvD.110l3029C}
{Cabrera}, T., {Palmese}, A., {Hu}, L., {et~al.} 2024, \bibinfo{title}{{Searching for electromagnetic emission in an AGN from the gravitational wave binary black hole merger candidate S230922g},} \prd, 110, 123029, \dodoi{10.1103/PhysRevD.110.123029}

\bibitem[{J. {Calder{\'o}n Bustillo} {et~al.}(2021){Calder{\'o}n Bustillo}, {Leong}, {Chandra}, {McKernan}, \& {Ford}}]{2021arXiv211212481C}
{Calder{\'o}n Bustillo}, J., {Leong}, S. H.~W., {Chandra}, K., {McKernan}, B., \& {Ford}, K.~E.~S. 2021, \bibinfo{title}{{GW190521 as a black-hole merger coincident with the ZTF19abanrhr flare},} arXiv e-prints, arXiv:2112.12481, \dodoi{10.48550/arXiv.2112.12481}

\bibitem[{H.-Y. {Chen} {et~al.}(2018){Chen}, {Fishbach}, \& {Holz}}]{2018Natur.562..545C}
{Chen}, H.-Y., {Fishbach}, M., \& {Holz}, D.~E. 2018, \bibinfo{title}{{A two per cent Hubble constant measurement from standard sirens within five years},} \nat, 562, 545, \dodoi{10.1038/s41586-018-0606-0}

\bibitem[{H.-Y. {Chen} {et~al.}(2022){Chen}, {Haster}, {Vitale}, {Farr}, \& {Isi}}]{2022MNRAS.513.2152C}
{Chen}, H.-Y., {Haster}, C.-J., {Vitale}, S., {Farr}, W.~M., \& {Isi}, M. 2022, \bibinfo{title}{{A standard siren cosmological measurement from the potential GW190521 electromagnetic counterpart ZTF19abanrhr},} \mnras, 513, 2152, \dodoi{10.1093/mnras/stac989}

\bibitem[{K. {Chen} \& Z.-G. {Dai}(2024){Chen} \& {Dai}}]{2024ApJ...961..206C}
{Chen}, K., \& {Dai}, Z.-G. 2024, \bibinfo{title}{{Electromagnetic Counterparts Powered by Kicked Remnants of Black Hole Binary Mergers in AGN Disks},} \apj, 961, 206, \dodoi{10.3847/1538-4357/ad0dfd}

\bibitem[{S. {Chen} \& K. {Jani}(2024){Chen} \& {Jani}}]{2024arXiv241102778C}
{Chen}, S., \& {Jani}, K. 2024, \bibinfo{title}{{Distinguishing the Demographics of Compact Binaries with Merger Entropy Index},} arXiv e-prints, arXiv:2411.02778, \dodoi{10.48550/arXiv.2411.02778}

\bibitem[{X. {Chen} \& Z. {Zhang}(2022){Chen} \& {Zhang}}]{2022PhRvD.106j3040C}
{Chen}, X., \& {Zhang}, Z. 2022, \bibinfo{title}{{Binaries wandering around supermassive black holes due to gravitoelectromagnetism},} \prd, 106, 103040, \dodoi{10.1103/PhysRevD.106.103040}

\bibitem[{B. {Cousins} {et~al.}(2025){Cousins}, {Schumacher}, {Ka-Wai Chung}, {Talbot}, {Callister}, {Holz}, \& {Yunes}}]{2025arXiv250301997C}
{Cousins}, B., {Schumacher}, K., {Ka-Wai Chung}, A., {et~al.} 2025, \bibinfo{title}{{The Stochastic Siren: Astrophysical Gravitational-Wave Background Measurements of the Hubble Constant},} arXiv e-prints, arXiv:2503.01997, \dodoi{10.48550/arXiv.2503.01997}

\bibitem[{W. {Del Pozzo}(2012){Del Pozzo}}]{2012PhRvD..86d3011D}
{Del Pozzo}, W. 2012, \bibinfo{title}{{Inference of cosmological parameters from gravitational waves: Applications to second generation interferometers},} \prd, 86, 043011, \dodoi{10.1103/PhysRevD.86.043011}

\bibitem[{V. {Delfavero} {et~al.}(2024){Delfavero}, {Ford}, {McKernan}, {Cook}, {Nathaniel}, {Postiglione}, {Ray}, \& {O'Shaughnessy}}]{2024arXiv241018815D}
{Delfavero}, V., {Ford}, K.~E.~S., {McKernan}, B., {et~al.} 2024, \bibinfo{title}{{McFacts III: Compact binary mergers from AGN disks over an entire synthetic universe},} arXiv e-prints, arXiv:2410.18815, \dodoi{10.48550/arXiv.2410.18815}

\bibitem[{E. {Di Valentino} {et~al.}(2021){Di Valentino}, {Mena}, {Pan}, {Visinelli}, {Yang}, {Melchiorri}, {Mota}, {Riess}, \& {Silk}}]{2021CQGra..38o3001D}
{Di Valentino}, E., {Mena}, O., {Pan}, S., {et~al.} 2021, \bibinfo{title}{{In the realm of the Hubble tension-a review of solutions},} Classical and Quantum Gravity, 38, 153001, \dodoi{10.1088/1361-6382/ac086d}

\bibitem[{G. {Fabj} {et~al.}(2020){Fabj}, {Nasim}, {Caban}, {Ford}, {McKernan}, \& {Bellovary}}]{2020MNRAS.499.2608F}
{Fabj}, G., {Nasim}, S.~S., {Caban}, F., {et~al.} 2020, \bibinfo{title}{{Aligning Nuclear Cluster Orbits with an Active Galactic Nucleus Accretion Disc},} \mnras, 499, 2608, \dodoi{10.1093/mnras/staa3004}

\bibitem[{X. {Fan} {et~al.}(2014){Fan}, {Messenger}, \& {Heng}}]{2014ApJ...795...43F}
{Fan}, X., {Messenger}, C., \& {Heng}, I.~S. 2014, \bibinfo{title}{{A Bayesian Approach to Multi-messenger Astronomy: Identification of Gravitational-wave Host Galaxies},} \apj, 795, 43, \dodoi{10.1088/0004-637X/795/1/43}

\bibitem[{X. {Fan} {et~al.}(2017){Fan}, {Messenger}, \& {Heng}}]{2017PhRvL.119r1102F}
{Fan}, X., {Messenger}, C., \& {Heng}, I.~S. 2017, \bibinfo{title}{{Probing Intrinsic Properties of Short Gamma-Ray Bursts with Gravitational Waves},} \prl, 119, 181102, \dodoi{10.1103/PhysRevLett.119.181102}

\bibitem[{Y. {Fang} \& Q.-G. {Huang}(2019){Fang} \& {Huang}}]{2019PhRvD..99j3005F}
{Fang}, Y., \& {Huang}, Q.-G. 2019, \bibinfo{title}{{Secular evolution of compact binaries revolving around a spinning massive black hole},} \prd, 99, 103005, \dodoi{10.1103/PhysRevD.99.103005}

\bibitem[{M. {Fishbach} {et~al.}(2017){Fishbach}, {Holz}, \& {Farr}}]{2017ApJ...840L..24F}
{Fishbach}, M., {Holz}, D.~E., \& {Farr}, B. 2017, \bibinfo{title}{{Are LIGO's Black Holes Made from Smaller Black Holes?},} \apjl, 840, L24, \dodoi{10.3847/2041-8213/aa7045}

\bibitem[{M. {Fishbach} {et~al.}(2019){Fishbach}, {Gray}, {Maga{\~n}a Hernandez}, {Qi}, {Sur}, {Acernese}, {Aiello}, {Allocca}, {Aloy}, {Amato}, {Antier}, {Ar{\`e}ne}, {Arnaud}, {Ascenzi}, {Astone}, {Aubin}, {Babak}, {Bacon}, {Badaracco}, {Bader}, {Baldaccini}, {Ballardin}, {Barone}, {Barsuglia}, {Barta}, {Basti}, {Bawaj}, {Bazzan}, {Bejger}, {Belahcene}, {Bernuzzi}, {Bersanetti}, {Bertolini}, {Bitossi}, {Bizouard}, {Blair}, {Bloemen}, {Boer}, {Bogaert}, {Bondu}, {Bonnand}, {Boom}, {Boschi}, {Bouffanais}, {Bozzi}, {Bradaschia}, {Brady}, {Branchesi}, {Briant}, {Brighenti}, {Brillet}, {Brisson}, {Bulik}, {Bulten}, {Buskulic}, {Buy}, {Cagnoli}, {Calloni}, {Canepa}, {Capocasa}, {Carbognani}, {Carullo}, {Casanueva Diaz}, {Casentini}, {Caudill}, {Cavalier}, {Cavalieri}, {Cella}, {Cerd{\'a}-Dur{\'a}n}, {Cerretani}, {Cesarini}, {Chaibi}, {Chassande-Mottin}, {Chatziioannou}, {Chen}, {Chincarini}, {Chiummo}, {Christensen}, {Chua}, {Ciani}, {Ciolfi}, {Cipriano}, {Cirone}, {Cleva}, {Coccia}, {Cohadon}, {Cohen}, {Conti},
  {Cordero-Carri{\'o}n}, {Cortese}, {Coughlin}, {Coulon}, {Croquette}, {Cuoco}, {D{\'a}lya}, {D'Antonio}, {Datrier}, {Dattilo}, {Davier}, {Degallaix}, {De Laurentis}, {Del{\'e}glise}, {Del Pozzo}, {Denys}, {De Pietri}, {De Rosa}, {De Rossi}, {DeSalvo}, {Dietrich}, {Di Fiore}, {Di Giovanni}, {Di Girolamo}, {Di Lieto}, {Di Pace}, {Di Palma}, {Di Renzo}, {Doctor}, {Drago}, {Ducoin}, {Eisenmann}, {Essick}, {Estevez}, {Fafone}, {Farinon}, {Farr}, {Feng}, {Ferrante}, {Ferrini}, {Fidecaro}, {Fiori}, {Fiorucci}, {Flaminio}, {Font}, {Fournier}, {Frasca}, {Frasconi}, {Frey}, {Gair}, {Gammaitoni}, {Garufi}, {Gemme}, {Genin}, {Gennai}, {George}, {Germain}, {Ghosh}, {Giacomazzo}, {Giazotto}, {Giordano}, {Gonzalez Castro}, {Gosselin}, {Gouaty}, {Grado}, {Granata}, {Greco}, {Groot}, {Gruning}, {Guidi}, {Guo}, {Halim}, {Harms}, {Haster}, {Heidmann}, {Heitmann}, {Hello}, {Hemming}, {Hendry}, {Hinderer}, {Hoak}, {Hofman}, {Holz}, {Hreibi}, {Huet}, {Idzkowski}, {Iess}, {Intini}, {Isac}, {Jacqmin}, {Jaranowski}, {Jonker},
  {Katsanevas}, {Katsavounidis}, {K{\'e}f{\'e}lian}, {Khan}, {Koekoek}, {Koley}, {Kowalska}, {Kr{\'o}lak}, {Kutynia}, {Lange}, {Lartaux-Vollard}, {Lazzaro}, {Leaci}, {Letendre}, {Li}, {Linde}, {Longo}, {Lorenzini}, {Loriette}, \& {Losurdo}}]{2019ApJ...871L..13F}
{Fishbach}, M., {Gray}, R., {Maga{\~n}a Hernandez}, I., {et~al.} 2019, \bibinfo{title}{{A Standard Siren Measurement of the Hubble Constant from GW170817 without the Electromagnetic Counterpart},} \apjl, 871, L13, \dodoi{10.3847/2041-8213/aaf96e}

\bibitem[{D. {Foreman-Mackey}(2016){Foreman-Mackey}}]{2016JOSS....1...24F}
{Foreman-Mackey}, D. 2016, \bibinfo{title}{{corner.py: Scatterplot matrices in Python},} The Journal of Open Source Software, 1, 24, \dodoi{10.21105/joss.00024}

\bibitem[{D. {Foreman-Mackey} {et~al.}(2013){Foreman-Mackey}, {Hogg}, {Lang}, \& {Goodman}}]{2013PASP..125..306F}
{Foreman-Mackey}, D., {Hogg}, D.~W., {Lang}, D., \& {Goodman}, J. 2013, \bibinfo{title}{{emcee: The MCMC Hammer},} \pasp, 125, 306, \dodoi{10.1086/670067}

\bibitem[{G. {Fragione} {et~al.}(2020){Fragione}, {Loeb}, \& {Rasio}}]{2020ApJ...902L..26F}
{Fragione}, G., {Loeb}, A., \& {Rasio}, F.~A. 2020, \bibinfo{title}{{On the Origin of GW190521-like Events from Repeated Black Hole Mergers in Star Clusters},} \apjl, 902, L26, \dodoi{10.3847/2041-8213/abbc0a}

\bibitem[{W.~L. {Freedman} {et~al.}(2019){Freedman}, {Madore}, {Hatt}, {Hoyt}, {Jang}, {Beaton}, {Burns}, {Lee}, {Monson}, {Neeley}, {Phillips}, {Rich}, \& {Seibert}}]{2019ApJ...882...34F}
{Freedman}, W.~L., {Madore}, B.~F., {Hatt}, D., {et~al.} 2019, \bibinfo{title}{{The Carnegie-Chicago Hubble Program. VIII. An Independent Determination of the Hubble Constant Based on the Tip of the Red Giant Branch},} \apj, 882, 34, \dodoi{10.3847/1538-4357/ab2f73}

\bibitem[{R. {Gamba} {et~al.}(2023){Gamba}, {Breschi}, {Carullo}, {Albanesi}, {Rettegno}, {Bernuzzi}, \& {Nagar}}]{2023NatAs...7...11G}
{Gamba}, R., {Breschi}, M., {Carullo}, G., {et~al.} 2023, \bibinfo{title}{{GW190521 as a dynamical capture of two nonspinning black holes},} Nature Astronomy, 7, 11, \dodoi{10.1038/s41550-022-01813-w}

\bibitem[{V. {Gayathri} {et~al.}(2021){Gayathri}, {Healy}, {Lange}, {O'Brien}, {Szczepanczyk}, {Bartos}, {Campanelli}, {Klimenko}, {Lousto}, \& {O'Shaughnessy}}]{2021ApJ...908L..34G}
{Gayathri}, V., {Healy}, J., {Lange}, J., {et~al.} 2021, \bibinfo{title}{{Measuring the Hubble Constant with GW190521 as an Eccentric black hole Merger and Its Potential Electromagnetic Counterpart},} \apjl, 908, L34, \dodoi{10.3847/2041-8213/abe388}

\bibitem[{V. {Gayathri} {et~al.}(2022){Gayathri}, {Healy}, {Lange}, {O'Brien}, {Szczepa{\'n}czyk}, {Bartos}, {Campanelli}, {Klimenko}, {Lousto}, \& {O'Shaughnessy}}]{2022NatAs...6..344G}
{Gayathri}, V., {Healy}, J., {Lange}, J., {et~al.} 2022, \bibinfo{title}{{Eccentricity estimate for black hole mergers with numerical relativity simulations},} Nature Astronomy, 6, 344, \dodoi{10.1038/s41550-021-01568-w}

\bibitem[{D. {Gerosa} \& E. {Berti}(2017){Gerosa} \& {Berti}}]{2017PhRvD..95l4046G}
{Gerosa}, D., \& {Berti}, E. 2017, \bibinfo{title}{{Are merging black holes born from stellar collapse or previous mergers?},} \prd, 95, 124046, \dodoi{10.1103/PhysRevD.95.124046}

\bibitem[{D. {Gerosa} \& M. {Fishbach}(2021){Gerosa} \& {Fishbach}}]{2021NatAs...5..749G}
{Gerosa}, D., \& {Fishbach}, M. 2021, \bibinfo{title}{{Hierarchical mergers of stellar-mass black holes and their gravitational-wave signatures},} Nature Astronomy, 5, 749, \dodoi{10.1038/s41550-021-01398-w}

\bibitem[{M.~J. {Graham} {et~al.}(2020){Graham}, {Ford}, {McKernan}, {Ross}, {Stern}, {Burdge}, {Coughlin}, {Djorgovski}, {Drake}, {Duev}, {Kasliwal}, {Mahabal}, {van Velzen}, {Belecki}, {Bellm}, {Burruss}, {Cenko}, {Cunningham}, {Helou}, {Kulkarni}, {Masci}, {Prince}, {Reiley}, {Rodriguez}, {Rusholme}, {Smith}, \& {Soumagnac}}]{2020PhRvL.124y1102G}
{Graham}, M.~J., {Ford}, K.~E.~S., {McKernan}, B., {et~al.} 2020, \bibinfo{title}{{Candidate Electromagnetic Counterpart to the Binary Black Hole Merger Gravitational-Wave Event S190521g$^{*}$},} \prl, 124, 251102, \dodoi{10.1103/PhysRevLett.124.251102}

\bibitem[{M.~J. {Graham} {et~al.}(2023){Graham}, {McKernan}, {Ford}, {Stern}, {Djorgovski}, {Coughlin}, {Burdge}, {Bellm}, {Helou}, {Mahabal}, {Masci}, {Purdum}, {Rosnet}, \& {Rusholme}}]{2023ApJ...942...99G}
{Graham}, M.~J., {McKernan}, B., {Ford}, K.~E.~S., {et~al.} 2023, \bibinfo{title}{{A Light in the Dark: Searching for Electromagnetic Counterparts to Black Hole-Black Hole Mergers in LIGO/Virgo O3 with the Zwicky Transient Facility},} \apj, 942, 99, \dodoi{10.3847/1538-4357/aca480}

\bibitem[{R. {Gray} {et~al.}(2020){Gray}, {Hernandez}, {Qi}, {Sur}, {Brady}, {Chen}, {Farr}, {Fishbach}, {Gair}, {Ghosh}, {Holz}, {Mastrogiovanni}, {Messenger}, {Steer}, \& {Veitch}}]{2020PhRvD.101l2001G}
{Gray}, R., {Hernandez}, I.~M., {Qi}, H., {et~al.} 2020, \bibinfo{title}{{Cosmological inference using gravitational wave standard sirens: A mock data analysis},} \prd, 101, 122001, \dodoi{10.1103/PhysRevD.101.122001}

\bibitem[{W.-B. {Han} {et~al.}(2024){Han}, {Yang}, {Tagawa}, {Jiang}, {Shen}, {Yun}, {Zhang}, \& {Zhong}}]{2024arXiv240101743H}
{Han}, W.-B., {Yang}, S.-C., {Tagawa}, H., {et~al.} 2024, \bibinfo{title}{{Indication for a compact object next to a LIGO-Virgo binary black hole merger},} arXiv e-prints, arXiv:2401.01743, \dodoi{10.48550/arXiv.2401.01743}

\bibitem[{C.~R. Harris {et~al.}(2020)Harris, Millman, van~der Walt, Gommers, Virtanen, Cournapeau, Wieser, Taylor, Berg, Smith, Kern, Picus, Hoyer, van Kerkwijk, Brett, Haldane, del R{\'{i}}o, Wiebe, Peterson, G{\'{e}}rard-Marchant, Sheppard, Reddy, Weckesser, Abbasi, Gohlke, \& Oliphant}]{harris2020array}
Harris, C.~R., Millman, K.~J., van~der Walt, S.~J., {et~al.} 2020, \bibinfo{title}{Array programming with {NumPy},} Nature, 585, 357, \dodoi{10.1038/s41586-020-2649-2}

\bibitem[{D.~E. {Holz} \& S.~A. {Hughes}(2005){Holz} \& {Hughes}}]{2005ApJ...629...15H}
{Holz}, D.~E., \& {Hughes}, S.~A. 2005, \bibinfo{title}{{Using Gravitational-Wave Standard Sirens},} \apj, 629, 15, \dodoi{10.1086/431341}

\bibitem[{J.~D. {Hunter}(2007){Hunter}}]{2007CSE.....9...90H}
{Hunter}, J.~D. 2007, \bibinfo{title}{{Matplotlib: A 2D Graphics Environment},} Computing in Science and Engineering, 9, 90, \dodoi{10.1109/MCSE.2007.55}

\bibitem[{S.-J. {Jin} {et~al.}(2023){Jin}, {Xing}, {Shao}, {Zhang}, \& {Zhang}}]{2023ChPhC..47f5104J}
{Jin}, S.-J., {Xing}, S.-S., {Shao}, Y., {Zhang}, J.-F., \& {Zhang}, X. 2023, \bibinfo{title}{{Joint constraints on cosmological parameters using future multi-band gravitational wave standard siren observations},} Chinese Physics C, 47, 065104, \dodoi{10.1088/1674-1137/acc8be}

\bibitem[{S.-J. {Jin} {et~al.}(2024){Jin}, {Zhang}, {Song}, {Zhang}, \& {Zhang}}]{2024SCPMA..6720412J}
{Jin}, S.-J., {Zhang}, Y.-Z., {Song}, J.-Y., {Zhang}, J.-F., \& {Zhang}, X. 2024, \bibinfo{title}{{Taiji-TianQin-LISA network: Precisely measuring the Hubble constant using both bright and dark sirens},} Science China Physics, Mechanics, and Astronomy, 67, 220412, \dodoi{10.1007/s11433-023-2276-1}

\bibitem[{M. {Kamionkowski} \& A.~G. {Riess}(2023){Kamionkowski} \& {Riess}}]{2023ARNPS..73..153K}
{Kamionkowski}, M., \& {Riess}, A.~G. 2023, \bibinfo{title}{{The Hubble Tension and Early Dark Energy},} Annual Review of Nuclear and Particle Science, 73, 153, \dodoi{10.1146/annurev-nucl-111422-024107}

\bibitem[{C. {Kimball} {et~al.}(2021){Kimball}, {Talbot}, {Berry}, {Zevin}, {Thrane}, {Kalogera}, {Buscicchio}, {Carney}, {Dent}, {Middleton}, {Payne}, {Veitch}, \& {Williams}}]{2021ApJ...915L..35K}
{Kimball}, C., {Talbot}, C., {Berry}, C. P.~L., {et~al.} 2021, \bibinfo{title}{{Evidence for Hierarchical Black Hole Mergers in the Second LIGO-Virgo Gravitational Wave Catalog},} \apjl, 915, L35, \dodoi{10.3847/2041-8213/ac0aef}

\bibitem[{J.-W. {Lee}(2025){Lee}}]{2025arXiv250211568L}
{Lee}, J.-W. 2025, \bibinfo{title}{{A solution to the Hubble tension with self-interacting ultralight dark matter},} arXiv e-prints, arXiv:2502.11568, \dodoi{10.48550/arXiv.2502.11568}

\bibitem[{G.-P. {Li}(2022{\natexlab{a}}){Li}}]{2022A&A...666A.194L}
{Li}, G.-P. 2022{\natexlab{a}}, \bibinfo{title}{{Constraining hierarchical mergers of binary black holes detectable with LIGO-Virgo},} \aap, 666, A194, \dodoi{10.1051/0004-6361/202244257}

\bibitem[{G.-P. {Li}(2022{\natexlab{b}}){Li}}]{2022PhRvD.105f3006L}
{Li}, G.-P. 2022{\natexlab{b}}, \bibinfo{title}{{Time-dependent stellar-mass binary black hole mergers in AGN disks: Mass distribution of hierarchical mergers},} \prd, 105, 063006, \dodoi{10.1103/PhysRevD.105.063006}

\bibitem[{G.-P. {Li} \& X.-L. {Fan}(2025{\natexlab{a}}){Li} \& {Fan}}]{2025ApJ...981..177L}
{Li}, G.-P., \& {Fan}, X.-L. 2025{\natexlab{a}}, \bibinfo{title}{{The Origin Channels of Hierarchical Binary Black Hole Mergers in the LIGO{\textendash}Virgo{\textendash}KAGRA O1, O2, and O3 Runs},} \apj, 981, 177, \dodoi{10.3847/1538-4357/adb578}

\bibitem[{G.-P. {Li} \& X.-L. {Fan}(2025{\natexlab{b}}){Li} \& {Fan}}]{PhysRevD.111.103016}
{Li}, G.-P., \& {Fan}, X.-L. 2025{\natexlab{b}}, \bibinfo{title}{Multimessenger hierarchical triple merger gravitational-wave event pair GW190514-GW190521 inside AGN $\mathrm{J}124942.3+344929$,} Phys. Rev. D, 111, 103016, \dodoi{10.1103/PhysRevD.111.103016}

\bibitem[{G.-P. {Li} \& X.-L. {Fan}(2025{\natexlab{c}}){Li} \& {Fan}}]{2025ApJ...984...63L}
{Li}, G.-P., \& {Fan}, X.-L. 2025{\natexlab{c}}, \bibinfo{title}{{The Nature of Gravitational Wave Events with Host Environment Escape Velocities},} \apj, 984, 63, \dodoi{10.3847/1538-4357/adc7bb}

\bibitem[{G.-P. {Li} {et~al.}(2023){Li}, {Lin}, \& {Yuan}}]{2023PhRvD.107f3007L}
{Li}, G.-P., {Lin}, D.-B., \& {Yuan}, Y. 2023, \bibinfo{title}{{Comparing hierarchical black hole mergers in star clusters and active galactic nuclei},} \prd, 107, 063007, \dodoi{10.1103/PhysRevD.107.063007}

\bibitem[{Y.-J. {Li} {et~al.}(2024{\natexlab{a}}){Li}, {Tang}, {Wang}, \& {Fan}}]{2024ApJ...976..153L}
{Li}, Y.-J., {Tang}, S.-P., {Wang}, Y.-Z., \& {Fan}, Y.-Z. 2024{\natexlab{a}}, \bibinfo{title}{{Multispectral Sirens: Gravitational-wave Cosmology with (Multi-) Subpopulations of Binary Black Holes},} \apj, 976, 153, \dodoi{10.3847/1538-4357/ad888b}

\bibitem[{Y.-J. {Li} {et~al.}(2024{\natexlab{b}}){Li}, {Wang}, {Tang}, \& {Fan}}]{2024PhRvL.133e1401L}
{Li}, Y.-J., {Wang}, Y.-Z., {Tang}, S.-P., \& {Fan}, Y.-Z. 2024{\natexlab{b}}, \bibinfo{title}{{Resolving the Stellar-Collapse and Hierarchical-Merger Origins of the Coalescing Black Holes},} \prl, 133, 051401, \dodoi{10.1103/PhysRevLett.133.051401}

\bibitem[{Y.-P. {Li} {et~al.}(2022){Li}, {Chen}, {Lin}, \& {Wang}}]{2022ApJ...928L...1L}
{Li}, Y.-P., {Chen}, Y.-X., {Lin}, D. N.~C., \& {Wang}, Z. 2022, \bibinfo{title}{{Spin Evolution of Stellar-mass Black Holes Embedded in AGN Disks: Orbital Eccentricity Produces Retrograde Circumstellar Flows},} \apjl, 928, L1, \dodoi{10.3847/2041-8213/ac5b61}

\bibitem[{ {LIGO Scientific Collaboration} {et~al.}(2015){LIGO Scientific Collaboration}, {Aasi}, {Abbott}, {Abbott}, {Abbott}, {Abernathy}, {Ackley}, {Adams}, {Adams}, {Addesso}, \& et~al.}]{2015CQGra..32g4001L}
{LIGO Scientific Collaboration}, {Aasi}, J., {Abbott}, B.~P., {et~al.} 2015, \bibinfo{title}{{Advanced LIGO},} Classical and Quantum Gravity, 32, 074001, \dodoi{10.1088/0264-9381/32/7/074001}

\bibitem[{B. {Liu} \& D. {Lai}(2021){Liu} \& {Lai}}]{2021MNRAS.502.2049L}
{Liu}, B., \& {Lai}, D. 2021, \bibinfo{title}{{Hierarchical black hole mergers in multiple systems: constrain the formation of GW190412-, GW190814-, and GW190521-like events},} \mnras, 502, 2049, \dodoi{10.1093/mnras/stab178}

\bibitem[{B. {Liu} {et~al.}(2019){Liu}, {Lai}, \& {Wang}}]{2019ApJ...883L...7L}
{Liu}, B., {Lai}, D., \& {Wang}, Y.-H. 2019, \bibinfo{title}{{Binary Mergers near a Supermassive Black Hole: Relativistic Effects in Triples},} \apjl, 883, L7, \dodoi{10.3847/2041-8213/ab40c0}

\bibitem[{Y. {Luo} {et~al.}(2023){Luo}, {Wu}, {Zhang}, {Wang}, {Ho}, \& {Yuan}}]{2023MNRAS.524.6015L}
{Luo}, Y., {Wu}, X.-J., {Zhang}, S.-R., {et~al.} 2023, \bibinfo{title}{{White dwarf-white dwarf collisions in AGN discs via close encounters},} \mnras, 524, 6015, \dodoi{10.1093/mnras/stad2188}

\bibitem[{E. {Macaulay} {et~al.}(2019){Macaulay}, {Nichol}, {Bacon}, {Brout}, {Davis}, {Zhang}, {Bassett}, {Scolnic}, {M{\"o}ller}, {D'Andrea}, {Hinton}, {Kessler}, {Kim}, {Lasker}, {Lidman}, {Sako}, {Smith}, {Sullivan}, {Abbott}, {Allam}, {Annis}, {Asorey}, {Avila}, {Bechtol}, {Brooks}, {Brown}, {Burke}, {Calcino}, {Carnero Rosell}, {Carollo}, {Carrasco Kind}, {Carretero}, {Castander}, {Collett}, {Crocce}, {Cunha}, {da Costa}, {Davis}, {De Vicente}, {Diehl}, {Doel}, {Drlica-Wagner}, {Eifler}, {Estrada}, {Evrard}, {Filippenko}, {Finley}, {Flaugher}, {Foley}, {Fosalba}, {Frieman}, {Galbany}, {Garc{\'\i}a-Bellido}, {Gaztanaga}, {Glazebrook}, {Gonz{\'a}lez-Gait{\'a}n}, {Gruen}, {Gruendl}, {Gschwend}, {Gutierrez}, {Hartley}, {Hollowood}, {Honscheid}, {Hoormann}, {Hoyle}, {Huterer}, {Jain}, {James}, {Jeltema}, {Kasai}, {Krause}, {Kuehn}, {Kuropatkin}, {Lahav}, {Lewis}, {Li}, {Lima}, {Lin}, {Maia}, {Marshall}, {Martini}, {Miquel}, {Nugent}, {Palmese}, {Pan}, {Plazas}, {Romer}, {Roodman}, {Sanchez}, {Scarpine},
  {Schindler}, {Schubnell}, {Serrano}, {Sevilla-Noarbe}, {Sharp}, {Soares-Santos}, {Sobreira}, {Sommer}, {Suchyta}, {Swann}, {Swanson}, {Tarle}, {Thomas}, {Thomas}, {Tucker}, {Uddin}, {Vikram}, {Walker}, {Wiseman}, \& {DES Collaboration}}]{2019MNRAS.486.2184M}
{Macaulay}, E., {Nichol}, R.~C., {Bacon}, D., {et~al.} 2019, \bibinfo{title}{{First cosmological results using Type Ia supernovae from the Dark Energy Survey: measurement of the Hubble constant},} \mnras, 486, 2184, \dodoi{10.1093/mnras/stz978}

\bibitem[{P. {Mahapatra} {et~al.}(2024){Mahapatra}, {Chattopadhyay}, {Gupta}, {Antonini}, {Favata}, {Sathyaprakash}, \& {Arun}}]{2024ApJ...975..117M}
{Mahapatra}, P., {Chattopadhyay}, D., {Gupta}, A., {et~al.} 2024, \bibinfo{title}{{Reconstructing the Genealogy of LIGO-Virgo Black Holes},} \apj, 975, 117, \dodoi{10.3847/1538-4357/ad781b}

\bibitem[{M. {Mancarella} {et~al.}(2024){Mancarella}, {Iacovelli}, {Foffa}, {Muttoni}, \& {Maggiore}}]{2024PhRvL.133z1001M}
{Mancarella}, M., {Iacovelli}, F., {Foffa}, S., {Muttoni}, N., \& {Maggiore}, M. 2024, \bibinfo{title}{{Accurate Standard Siren Cosmology with Joint Gravitational-Wave and <inline-formula><mml:math display=``inline''><mml:mi>{\ensuremath{\gamma}}</mml:mi></mml:math></inline-formula>-Ray Burst Observations},} \prl, 133, 261001, \dodoi{10.1103/PhysRevLett.133.261001}

\bibitem[{B. {McKernan} {et~al.}(2020){McKernan}, {Ford}, {O'Shaugnessy}, \& {Wysocki}}]{2020MNRAS.494.1203M}
{McKernan}, B., {Ford}, K.~E.~S., {O'Shaugnessy}, R., \& {Wysocki}, D. 2020, \bibinfo{title}{{Monte Carlo simulations of black hole mergers in AGN discs: Low {\ensuremath{\chi}}$_{eff}$ mergers and predictions for LIGO},} \mnras, 494, 1203, \dodoi{10.1093/mnras/staa740}

\bibitem[{B. {McKernan} {et~al.}(2018){McKernan}, {Ford}, {Bellovary}, {Leigh}, {Haiman}, {Kocsis}, {Lyra}, {Mac Low}, {Metzger}, {O'Dowd}, {Endlich}, \& {Rosen}}]{2018ApJ...866...66M}
{McKernan}, B., {Ford}, K.~E.~S., {Bellovary}, J., {et~al.} 2018, \bibinfo{title}{{Constraining Stellar-mass Black Hole Mergers in AGN Disks Detectable with LIGO},} \apj, 866, 66, \dodoi{10.3847/1538-4357/aadae5}

\bibitem[{B. {McKernan} {et~al.}(2019){McKernan}, {Ford}, {Bartos}, {Graham}, {Lyra}, {Marka}, {Marka}, {Ross}, {Stern}, \& {Yang}}]{2019ApJ...884L..50M}
{McKernan}, B., {Ford}, K.~E.~S., {Bartos}, I., {et~al.} 2019, \bibinfo{title}{{Ram-pressure Stripping of a Kicked Hill Sphere: Prompt Electromagnetic Emission from the Merger of Stellar Mass Black Holes in an AGN Accretion Disk},} \apjl, 884, L50, \dodoi{10.3847/2041-8213/ab4886}

\bibitem[{S.~L. {Morton} {et~al.}(2023){Morton}, {Rinaldi}, {Torres-Orjuela}, {Derdzinski}, {Vaccaro}, \& {Del Pozzo}}]{2023PhRvD.108l3039M}
{Morton}, S.~L., {Rinaldi}, S., {Torres-Orjuela}, A., {et~al.} 2023, \bibinfo{title}{{GW190521: A binary black hole merger inside an active galactic nucleus?},} \prd, 108, 123039, \dodoi{10.1103/PhysRevD.108.123039}

\bibitem[{S. {Mukherjee} {et~al.}(2020){Mukherjee}, {Ghosh}, {Graham}, {Karathanasis}, {Kasliwal}, {Maga{\~n}a Hernandez}, {Nissanke}, {Silvestri}, \& {Wandelt}}]{2020arXiv200914199M}
{Mukherjee}, S., {Ghosh}, A., {Graham}, M.~J., {et~al.} 2020, \bibinfo{title}{{First measurement of the Hubble parameter from bright binary black hole GW190521},} arXiv e-prints, arXiv:2009.14199, \dodoi{10.48550/arXiv.2009.14199}

\bibitem[{F. {Niedermann} \& M.~S. {Sloth}(2020){Niedermann} \& {Sloth}}]{2020PhRvD.102f3527N}
{Niedermann}, F., \& {Sloth}, M.~S. 2020, \bibinfo{title}{{Resolving the Hubble tension with new early dark energy},} \prd, 102, 063527, \dodoi{10.1103/PhysRevD.102.063527}

\bibitem[{S. {Nissanke} {et~al.}(2013){Nissanke}, {Holz}, {Dalal}, {Hughes}, {Sievers}, \& {Hirata}}]{2013arXiv1307.2638N}
{Nissanke}, S., {Holz}, D.~E., {Dalal}, N., {et~al.} 2013, \bibinfo{title}{{Determining the Hubble constant from gravitational wave observations of merging compact binaries},} arXiv e-prints, arXiv:1307.2638, \dodoi{10.48550/arXiv.1307.2638}

\bibitem[{R.~M. {O'Leary} {et~al.}(2016){O'Leary}, {Meiron}, \& {Kocsis}}]{2016ApJ...824L..12O}
{O'Leary}, R.~M., {Meiron}, Y., \& {Kocsis}, B. 2016, \bibinfo{title}{{Dynamical Formation Signatures of Black Hole Binaries in the First Detected Mergers by LIGO},} \apjl, 824, L12, \dodoi{10.3847/2041-8205/824/1/L12}

\bibitem[{P. {Peng} \& X. {Chen}(2021){Peng} \& {Chen}}]{2021MNRAS.505.1324P}
{Peng}, P., \& {Chen}, X. 2021, \bibinfo{title}{{The last migration trap of compact objects in AGN accretion disc},} \mnras, 505, 1324, \dodoi{10.1093/mnras/stab1419}

\bibitem[{F. {Perez} \& B.~E. {Granger}(2007){Perez} \& {Granger}}]{2007CSE.....9c..21P}
{Perez}, F., \& {Granger}, B.~E. 2007, \bibinfo{title}{{IPython: A System for Interactive Scientific Computing},} Computing in Science and Engineering, 9, 21, \dodoi{10.1109/MCSE.2007.53}

\bibitem[{D.~W. {Pesce} {et~al.}(2020){Pesce}, {Braatz}, {Reid}, {Riess}, {Scolnic}, {Condon}, {Gao}, {Henkel}, {Impellizzeri}, {Kuo}, \& {Lo}}]{2020ApJ...891L...1P}
{Pesce}, D.~W., {Braatz}, J.~A., {Reid}, M.~J., {et~al.} 2020, \bibinfo{title}{{The Megamaser Cosmology Project. XIII. Combined Hubble Constant Constraints},} \apjl, 891, L1, \dodoi{10.3847/2041-8213/ab75f0}

\bibitem[{ {Planck Collaboration} {et~al.}(2016){Planck Collaboration}, {Ade}, {Aghanim}, {Arnaud}, {Ashdown}, {Aumont}, {Baccigalupi}, {Banday}, {Barreiro}, {Bartlett}, {Bartolo}, {Battaner}, {Battye}, {Benabed}, {Beno{\^\i}t}, {Benoit-L{\'e}vy}, {Bernard}, {Bersanelli}, {Bielewicz}, {Bock}, {Bonaldi}, {Bonavera}, {Bond}, {Borrill}, {Bouchet}, {Boulanger}, {Bucher}, {Burigana}, {Butler}, {Calabrese}, {Cardoso}, {Catalano}, {Challinor}, {Chamballu}, {Chary}, {Chiang}, {Chluba}, {Christensen}, {Church}, {Clements}, {Colombi}, {Colombo}, {Combet}, {Coulais}, {Crill}, {Curto}, {Cuttaia}, {Danese}, {Davies}, {Davis}, {de Bernardis}, {de Rosa}, {de Zotti}, {Delabrouille}, {D{\'e}sert}, {Di Valentino}, {Dickinson}, {Diego}, {Dolag}, {Dole}, {Donzelli}, {Dor{\'e}}, {Douspis}, {Ducout}, {Dunkley}, {Dupac}, {Efstathiou}, {Elsner}, {En{\ss}lin}, {Eriksen}, {Farhang}, {Fergusson}, {Finelli}, {Forni}, {Frailis}, {Fraisse}, {Franceschi}, {Frejsel}, {Galeotta}, {Galli}, {Ganga}, {Gauthier}, {Gerbino}, {Ghosh}, {Giard},
  {Giraud-H{\'e}raud}, {Giusarma}, {Gjerl{\o}w}, {Gonz{\'a}lez-Nuevo}, {G{\'o}rski}, {Gratton}, {Gregorio}, {Gruppuso}, {Gudmundsson}, {Hamann}, {Hansen}, {Hanson}, {Harrison}, {Helou}, {Henrot-Versill{\'e}}, {Hern{\'a}ndez-Monteagudo}, {Herranz}, {Hildebrandt}, {Hivon}, {Hobson}, {Holmes}, {Hornstrup}, {Hovest}, {Huang}, {Huffenberger}, {Hurier}, {Jaffe}, {Jaffe}, {Jones}, {Juvela}, {Keih{\"a}nen}, {Keskitalo}, {Kisner}, {Kneissl}, {Knoche}, {Knox}, {Kunz}, {Kurki-Suonio}, {Lagache}, {L{\"a}hteenm{\"a}ki}, {Lamarre}, {Lasenby}, {Lattanzi}, {Lawrence}, {Leahy}, {Leonardi}, {Lesgourgues}, {Levrier}, {Lewis}, {Liguori}, {Lilje}, {Linden-V{\o}rnle}, {L{\'o}pez-Caniego}, {Lubin}, {Mac{\'\i}as-P{\'e}rez}, {Maggio}, {Maino}, {Mandolesi}, {Mangilli}, {Marchini}, {Maris}, {Martin}, {Martinelli}, {Mart{\'\i}nez-Gonz{\'a}lez}, {Masi}, {Matarrese}, {McGehee}, {Meinhold}, {Melchiorri}, {Melin}, {Mendes}, {Mennella}, {Migliaccio}, {Millea}, {Mitra}, {Miville-Desch{\^e}nes}, {Moneti}, {Montier}, {Morgante}, {Mortlock},
  {Moss}, {Munshi}, {Murphy}, {Naselsky}, {Nati}, {Natoli}, {Netterfield}, {N{\o}rgaard-Nielsen}, {Noviello}, {Novikov}, {Novikov}, {Oxborrow}, {Paci}, {Pagano}, {Pajot}, {Paladini}, {Paoletti}, {Partridge}, {Pasian}, {Patanchon}, {Pearson}, {Perdereau}, {Perotto}, {Perrotta}, {Pettorino}, {Piacentini}, {Piat}, {Pierpaoli}, {Pietrobon}, {Plaszczynski}, {Pointecouteau}, {Polenta}, {Popa}, {Pratt}, \& {Pr{\'e}zeau}}]{2016A&A...594A..13P}
{Planck Collaboration}, {Ade}, P.~A.~R., {Aghanim}, N., {et~al.} 2016, \bibinfo{title}{{Planck 2015 results. XIII. Cosmological parameters},} \aap, 594, A13, \dodoi{10.1051/0004-6361/201525830}

\bibitem[{ {Planck Collaboration} {et~al.}(2020){Planck Collaboration}, {Aghanim}, {Akrami}, {Ashdown}, {Aumont}, {Baccigalupi}, {Ballardini}, {Banday}, {Barreiro}, {Bartolo}, {Basak}, {Battye}, {Benabed}, {Bernard}, {Bersanelli}, {Bielewicz}, {Bock}, {Bond}, {Borrill}, {Bouchet}, {Boulanger}, {Bucher}, {Burigana}, {Butler}, {Calabrese}, {Cardoso}, {Carron}, {Challinor}, {Chiang}, {Chluba}, {Colombo}, {Combet}, {Contreras}, {Crill}, {Cuttaia}, {de Bernardis}, {de Zotti}, {Delabrouille}, {Delouis}, {Di Valentino}, {Diego}, {Dor{\'e}}, {Douspis}, {Ducout}, {Dupac}, {Dusini}, {Efstathiou}, {Elsner}, {En{\ss}lin}, {Eriksen}, {Fantaye}, {Farhang}, {Fergusson}, {Fernandez-Cobos}, {Finelli}, {Forastieri}, {Frailis}, {Fraisse}, {Franceschi}, {Frolov}, {Galeotta}, {Galli}, {Ganga}, {G{\'e}nova-Santos}, {Gerbino}, {Ghosh}, {Gonz{\'a}lez-Nuevo}, {G{\'o}rski}, {Gratton}, {Gruppuso}, {Gudmundsson}, {Hamann}, {Handley}, {Hansen}, {Herranz}, {Hildebrandt}, {Hivon}, {Huang}, {Jaffe}, {Jones}, {Karakci}, {Keih{\"a}nen},
  {Keskitalo}, {Kiiveri}, {Kim}, {Kisner}, {Knox}, {Krachmalnicoff}, {Kunz}, {Kurki-Suonio}, {Lagache}, {Lamarre}, {Lasenby}, {Lattanzi}, {Lawrence}, {Le Jeune}, {Lemos}, {Lesgourgues}, {Levrier}, {Lewis}, {Liguori}, {Lilje}, {Lilley}, {Lindholm}, {L{\'o}pez-Caniego}, {Lubin}, {Ma}, {Mac{\'\i}as-P{\'e}rez}, {Maggio}, {Maino}, {Mandolesi}, {Mangilli}, {Marcos-Caballero}, {Maris}, {Martin}, {Martinelli}, {Mart{\'\i}nez-Gonz{\'a}lez}, {Matarrese}, {Mauri}, {McEwen}, {Meinhold}, {Melchiorri}, {Mennella}, {Migliaccio}, {Millea}, {Mitra}, {Miville-Desch{\^e}nes}, {Molinari}, {Montier}, {Morgante}, {Moss}, {Natoli}, {N{\o}rgaard-Nielsen}, {Pagano}, {Paoletti}, {Partridge}, {Patanchon}, {Peiris}, {Perrotta}, {Pettorino}, {Piacentini}, {Polastri}, {Polenta}, {Puget}, {Rachen}, {Reinecke}, {Remazeilles}, {Renzi}, {Rocha}, {Rosset}, {Roudier}, {Rubi{\~n}o-Mart{\'\i}n}, {Ruiz-Granados}, {Salvati}, {Sandri}, {Savelainen}, {Scott}, {Shellard}, {Sirignano}, {Sirri}, {Spencer}, {Sunyaev}, {Suur-Uski}, {Tauber}, {Tavagnacco},
  {Tenti}, {Toffolatti}, {Tomasi}, {Trombetti}, {Valenziano}, {Valiviita}, {Van Tent}, {Vibert}, {Vielva}, {Villa}, {Vittorio}, {Wandelt}, {Wehus}, {White}, {White}, {Zacchei}, \& {Zonca}}]{2020A&A...641A...6P}
{Planck Collaboration}, {Aghanim}, N., {Akrami}, Y., {et~al.} 2020, \bibinfo{title}{{Planck 2018 results. VI. Cosmological parameters},} \aap, 641, A6, \dodoi{10.1051/0004-6361/201833910}

\bibitem[{M. {Punturo} {et~al.}(2010{\natexlab{a}}){Punturo}, {Abernathy}, {Acernese}, {Allen}, {Andersson}, {Arun}, {Barone}, {Barr}, {Barsuglia}, {Beker}, {Beveridge}, {Birindelli}, {Bose}, {Bosi}, {Braccini}, {Bradaschia}, {Bulik}, {Calloni}, {Cella}, {Chassande Mottin}, {Chelkowski}, {Chincarini}, {Clark}, {Coccia}, {Colacino}, {Colas}, {Cumming}, {Cunningham}, {Cuoco}, {Danilishin}, {Danzmann}, {De Luca}, {De Salvo}, {Dent}, {De Rosa}, {Di Fiore}, {Di Virgilio}, {Doets}, {Fafone}, {Falferi}, {Flaminio}, {Franc}, {Frasconi}, {Freise}, {Fulda}, {Gair}, {Gemme}, {Gennai}, {Giazotto}, {Glampedakis}, {Granata}, {Grote}, {Guidi}, {Hammond}, {Hannam}, {Harms}, {Heinert}, {Hendry}, {Heng}, {Hennes}, {Hild}, {Hough}, {Husa}, {Huttner}, {Jones}, {Khalili}, {Kokeyama}, {Kokkotas}, {Krishnan}, {Lorenzini}, {L{\"u}ck}, {Majorana}, {Mandel}, {Mandic}, {Martin}, {Michel}, {Minenkov}, {Morgado}, {Mosca}, {Mours}, {M{\"u}ller{\textendash}Ebhardt}, {Murray}, {Nawrodt}, {Nelson}, {Oshaughnessy}, {Ott}, {Palomba}, {Paoli},
  {Parguez}, {Pasqualetti}, {Passaquieti}, {Passuello}, {Pinard}, {Poggiani}, {Popolizio}, {Prato}, {Puppo}, {Rabeling}, {Rapagnani}, {Read}, {Regimbau}, {Rehbein}, {Reid}, {Rezzolla}, {Ricci}, {Richard}, {Rocchi}, {Rowan}, {R{\"u}diger}, {Sassolas}, {Sathyaprakash}, {Schnabel}, {Schwarz}, {Seidel}, {Sintes}, {Somiya}, {Speirits}, {Strain}, {Strigin}, {Sutton}, {Tarabrin}, {Th{\"u}ring}, {van den Brand}, {van Leewen}, {van Veggel}, {van den Broeck}, {Vecchio}, {Veitch}, {Vetrano}, {Vicere}, {Vyatchanin}, {Willke}, {Woan}, {Wolfango}, \& {Yamamoto}}]{2010CQGra..27s4002P}
{Punturo}, M., {Abernathy}, M., {Acernese}, F., {et~al.} 2010{\natexlab{a}}, \bibinfo{title}{{The Einstein Telescope: a third-generation gravitational wave observatory},} Classical and Quantum Gravity, 27, 194002, \dodoi{10.1088/0264-9381/27/19/194002}

\bibitem[{M. {Punturo} {et~al.}(2010{\natexlab{b}}){Punturo}, {Abernathy}, {Acernese}, {Allen}, {Andersson}, {Arun}, {Barone}, {Barr}, {Barsuglia}, {Beker}, {Beveridge}, {Birindelli}, {Bose}, {Bosi}, {Braccini}, {Bradaschia}, {Bulik}, {Calloni}, {Cella}, {Chassande Mottin}, {Chelkowski}, {Chincarini}, {Clark}, {Coccia}, {Colacino}, {Colas}, {Cumming}, {Cunningham}, {Cuoco}, {Danilishin}, {Danzmann}, {De Luca}, {De Salvo}, {Dent}, {Derosa}, {Di Fiore}, {Di Virgilio}, {Doets}, {Fafone}, {Falferi}, {Flaminio}, {Franc}, {Frasconi}, {Freise}, {Fulda}, {Gair}, {Gemme}, {Gennai}, {Giazotto}, {Glampedakis}, {Granata}, {Grote}, {Guidi}, {Hammond}, {Hannam}, {Harms}, {Heinert}, {Hendry}, {Heng}, {Hennes}, {Hild}, {Hough}, {Husa}, {Huttner}, {Jones}, {Khalili}, {Kokeyama}, {Kokkotas}, {Krishnan}, {Lorenzini}, {L{\"u}ck}, {Majorana}, {Mandel}, {Mandic}, {Martin}, {Michel}, {Minenkov}, {Morgado}, {Mosca}, {Mours}, {M{\"u}ller-Ebhardt}, {Murray}, {Nawrodt}, {Nelson}, {Oshaughnessy}, {Ott}, {Palomba}, {Paoli}, {Parguez},
  {Pasqualetti}, {Passaquieti}, {Passuello}, {Pinard}, {Poggiani}, {Popolizio}, {Prato}, {Puppo}, {Rabeling}, {Rapagnani}, {Read}, {Regimbau}, {Rehbein}, {Reid}, {Rezzolla}, {Ricci}, {Richard}, {Rocchi}, {Rowan}, {R{\"u}diger}, {Sassolas}, {Sathyaprakash}, {Schnabel}, {Schwarz}, {Seidel}, {Sintes}, {Somiya}, {Speirits}, {Strain}, {Strigin}, {Sutton}, {Tarabrin}, {van den Brand}, {van Leewen}, {van Veggel}, {van den Broeck}, {Vecchio}, {Veitch}, {Vetrano}, {Vicere}, {Vyatchanin}, {Willke}, {Woan}, {Wolfango}, \& {Yamamoto}}]{2010CQGra..27h4007P}
{Punturo}, M., {Abernathy}, M., {Acernese}, F., {et~al.} 2010{\natexlab{b}}, \bibinfo{title}{{The third generation of gravitational wave observatories and their science reach},} Classical and Quantum Gravity, 27, 084007, \dodoi{10.1088/0264-9381/27/8/084007}

\bibitem[{A.~G. {Riess} {et~al.}(2019){Riess}, {Casertano}, {Yuan}, {Macri}, \& {Scolnic}}]{2019ApJ...876...85R}
{Riess}, A.~G., {Casertano}, S., {Yuan}, W., {Macri}, L.~M., \& {Scolnic}, D. 2019, \bibinfo{title}{{Large Magellanic Cloud Cepheid Standards Provide a 1\% Foundation for the Determination of the Hubble Constant and Stronger Evidence for Physics beyond {\ensuremath{\Lambda}}CDM},} \apj, 876, 85, \dodoi{10.3847/1538-4357/ab1422}

\bibitem[{A.~G. {Riess} {et~al.}(2016){Riess}, {Macri}, {Hoffmann}, {Scolnic}, {Casertano}, {Filippenko}, {Tucker}, {Reid}, {Jones}, {Silverman}, {Chornock}, {Challis}, {Yuan}, {Brown}, \& {Foley}}]{2016ApJ...826...56R}
{Riess}, A.~G., {Macri}, L.~M., {Hoffmann}, S.~L., {et~al.} 2016, \bibinfo{title}{{A 2.4\% Determination of the Local Value of the Hubble Constant},} \apj, 826, 56, \dodoi{10.3847/0004-637X/826/1/56}

\bibitem[{I. {Romero-Shaw} {et~al.}(2020){Romero-Shaw}, {Lasky}, {Thrane}, \& {Calder{\'o}n Bustillo}}]{2020ApJ...903L...5R}
{Romero-Shaw}, I., {Lasky}, P.~D., {Thrane}, E., \& {Calder{\'o}n Bustillo}, J. 2020, \bibinfo{title}{{GW190521: Orbital Eccentricity and Signatures of Dynamical Formation in a Binary Black Hole Merger Signal},} \apjl, 903, L5, \dodoi{10.3847/2041-8213/abbe26}

\bibitem[{J. {Sakstein} \& M. {Trodden}(2020){Sakstein} \& {Trodden}}]{2020PhRvL.124p1301S}
{Sakstein}, J., \& {Trodden}, M. 2020, \bibinfo{title}{{Early Dark Energy from Massive Neutrinos as a Natural Resolution of the Hubble Tension},} \prl, 124, 161301, \dodoi{10.1103/PhysRevLett.124.161301}

\bibitem[{J. {Samsing} \& T. {Ilan}(2018){Samsing} \& {Ilan}}]{2018MNRAS.476.1548S}
{Samsing}, J., \& {Ilan}, T. 2018, \bibinfo{title}{{Topology of black hole binary-single interactions},} \mnras, 476, 1548, \dodoi{10.1093/mnras/sty197}

\bibitem[{J. {Samsing} \& T. {Ilan}(2019){Samsing} \& {Ilan}}]{2019MNRAS.482...30S}
{Samsing}, J., \& {Ilan}, T. 2019, \bibinfo{title}{{Double gravitational wave mergers},} \mnras, 482, 30, \dodoi{10.1093/mnras/sty2249}

\bibitem[{J. {Samsing} {et~al.}(2022){Samsing}, {Bartos}, {D'Orazio}, {Haiman}, {Kocsis}, {Leigh}, {Liu}, {Pessah}, \& {Tagawa}}]{2022Natur.603..237S}
{Samsing}, J., {Bartos}, I., {D'Orazio}, D.~J., {et~al.} 2022, \bibinfo{title}{{AGN as potential factories for eccentric black hole mergers},} \nat, 603, 237, \dodoi{10.1038/s41586-021-04333-1}

\bibitem[{B.~S. {Sathyaprakash} {et~al.}(2010){Sathyaprakash}, {Schutz}, \& {Van Den Broeck}}]{2010CQGra..27u5006S}
{Sathyaprakash}, B.~S., {Schutz}, B.~F., \& {Van Den Broeck}, C. 2010, \bibinfo{title}{{Cosmography with the Einstein Telescope},} Classical and Quantum Gravity, 27, 215006, \dodoi{10.1088/0264-9381/27/21/215006}

\bibitem[{B.~F. {Schutz}(1986){Schutz}}]{1986Natur.323..310S}
{Schutz}, B.~F. 1986, \bibinfo{title}{{Determining the Hubble constant from gravitational wave observations},} \nat, 323, 310, \dodoi{10.1038/323310a0}

\bibitem[{K. {Somiya}(2012){Somiya}}]{2012CQGra..29l4007S}
{Somiya}, K. 2012, \bibinfo{title}{{Detector configuration of KAGRA-the Japanese cryogenic gravitational-wave detector},} Classical and Quantum Gravity, 29, 124007, \dodoi{10.1088/0264-9381/29/12/124007}

\bibitem[{T.~F.~M. {Spieksma} {et~al.}(2025){Spieksma}, {Cardoso}, {Carullo}, {Della Rocca}, \& {Duque}}]{2025PhRvL.134h1402S}
{Spieksma}, T. F.~M., {Cardoso}, V., {Carullo}, G., {Della Rocca}, M., \& {Duque}, F. 2025, \bibinfo{title}{{Black Hole Spectroscopy in Environments: Detectability Prospects},} \prl, 134, 081402, \dodoi{10.1103/PhysRevLett.134.081402}

\bibitem[{H. {Tagawa} {et~al.}(2020){Tagawa}, {Haiman}, \& {Kocsis}}]{2020ApJ...898...25T}
{Tagawa}, H., {Haiman}, Z., \& {Kocsis}, B. 2020, \bibinfo{title}{{Formation and Evolution of Compact-object Binaries in AGN Disks},} \apj, 898, 25, \dodoi{10.3847/1538-4357/ab9b8c}

\bibitem[{H. {Tagawa} {et~al.}(2023){Tagawa}, {Kimura}, \& {Haiman}}]{2023ApJ...955...23T}
{Tagawa}, H., {Kimura}, S.~S., \& {Haiman}, Z. 2023, \bibinfo{title}{{High-energy Electromagnetic, Neutrino, and Cosmic-Ray Emission by Stellar-mass Black Holes in Disks of Active Galactic Nuclei},} \apj, 955, 23, \dodoi{10.3847/1538-4357/ace71d}

\bibitem[{H. {Tong} {et~al.}(2025){Tong}, {Fishbach}, \& {Thrane}}]{2025arXiv250210780T}
{Tong}, H., {Fishbach}, M., \& {Thrane}, E. 2025, \bibinfo{title}{{Spinning spectral sirens: Robust cosmological measurement using mass-spin correlations in the binary black hole population},} arXiv e-prints, arXiv:2502.10780, \dodoi{10.48550/arXiv.2502.10780}

\bibitem[{A. {Torres-Orjuela} \& X. {Chen}(2023){Torres-Orjuela} \& {Chen}}]{2023PhRvD.107d3027T}
{Torres-Orjuela}, A., \& {Chen}, X. 2023, \bibinfo{title}{{Moving gravitational wave sources at cosmological distances: Impact on the measurement of the Hubble constant},} \prd, 107, 043027, \dodoi{10.1103/PhysRevD.107.043027}

\bibitem[{N. {Veronesi} {et~al.}(2023){Veronesi}, {Rossi}, \& {van Velzen}}]{2023MNRAS.526.6031V}
{Veronesi}, N., {Rossi}, E.~M., \& {van Velzen}, S. 2023, \bibinfo{title}{{The most luminous AGN do not produce the majority of the detected stellar-mass black hole binary mergers in the local Universe},} \mnras, 526, 6031, \dodoi{10.1093/mnras/stad3157}

\bibitem[{N. {Veronesi} {et~al.}(2022){Veronesi}, {Rossi}, {van Velzen}, \& {Buscicchio}}]{2022MNRAS.514.2092V}
{Veronesi}, N., {Rossi}, E.~M., {van Velzen}, S., \& {Buscicchio}, R. 2022, \bibinfo{title}{{Detectability of a spatial correlation between stellar mass black hole mergers and active galactic nuclei in the local Universe},} \mnras, 514, 2092, \dodoi{10.1093/mnras/stac1346}

\bibitem[{N. {Veronesi} {et~al.}(2024{\natexlab{a}}){Veronesi}, {van Velzen}, \& {Rossi}}]{2024arXiv240505318V}
{Veronesi}, N., {van Velzen}, S., \& {Rossi}, E.~M. 2024{\natexlab{a}}, \bibinfo{title}{{AGN flares as counterparts to the mergers detected by LIGO and Virgo: a novel spatial correlation analysis},} arXiv e-prints, arXiv:2405.05318, \dodoi{10.48550/arXiv.2405.05318}

\bibitem[{N. {Veronesi} {et~al.}(2024{\natexlab{b}}){Veronesi}, {van Velzen}, {Rossi}, \& {Storey-Fisher}}]{2024arXiv240721568V}
{Veronesi}, N., {van Velzen}, S., {Rossi}, E.~M., \& {Storey-Fisher}, K. 2024{\natexlab{b}}, \bibinfo{title}{{Constraining the AGN formation channel for detected black hole binary mergers up to z=1.5 with the Quaia catalogue},} arXiv e-prints, arXiv:2407.21568, \dodoi{10.48550/arXiv.2407.21568}

\bibitem[{D. {Veske} {et~al.}(2020){Veske}, {M{\'a}rka}, {Sullivan}, {Bartos}, {Rainer Corley}, {Samsing}, \& {M{\'a}rka}}]{2020MNRAS.498L..46V}
{Veske}, D., {M{\'a}rka}, Z., {Sullivan}, A.~G., {et~al.} 2020, \bibinfo{title}{{Have hierarchical three-body mergers been detected by LIGO/Virgo?},} \mnras, 498, L46, \dodoi{10.1093/mnrasl/slaa123}

\bibitem[{D. {Veske} {et~al.}(2021){Veske}, {Sullivan}, {M{\'a}rka}, {Bartos}, {Corley}, {Samsing}, {Buscicchio}, \& {M{\'a}rka}}]{2021ApJ...907L..48V}
{Veske}, D., {Sullivan}, A.~G., {M{\'a}rka}, Z., {et~al.} 2021, \bibinfo{title}{{Search for Black Hole Merger Families},} \apjl, 907, L48, \dodoi{10.3847/2041-8213/abd721}

\bibitem[{P. Virtanen {et~al.}(2020)Virtanen, Gommers, Oliphant, Haberland, Reddy, Cournapeau, Burovski, Peterson, Weckesser, Bright, {van der Walt}, Brett, Wilson, Millman, Mayorov, Nelson, Jones, Kern, Larson, Carey, Polat, Feng, Moore, {VanderPlas}, Laxalde, Perktold, Cimrman, Henriksen, Quintero, Harris, Archibald, Ribeiro, Pedregosa, {van Mulbregt}, \& {SciPy 1.0 Contributors}}]{2020SciPy-NMeth}
Virtanen, P., Gommers, R., Oliphant, T.~E., {et~al.} 2020, \bibinfo{title}{{{SciPy} 1.0: Fundamental Algorithms for Scientific Computing in Python},} Nature Methods, 17, 261, \dodoi{10.1038/s41592-019-0686-2}

\bibitem[{J.-M. {Wang} {et~al.}(2021){Wang}, {Liu}, {Ho}, {Li}, \& {Du}}]{2021ApJ...916L..17W}
{Wang}, J.-M., {Liu}, J.-R., {Ho}, L.~C., {Li}, Y.-R., \& {Du}, P. 2021, \bibinfo{title}{{Accretion-modified Stars in Accretion Disks of Active Galactic Nuclei: Gravitational-wave Bursts and Electromagnetic Counterparts from Merging Stellar Black Hole Binaries},} \apjl, 916, L17, \dodoi{10.3847/2041-8213/ac0b46}

\bibitem[{M. {Wang} {et~al.}(2025){Wang}, {Ma}, {Li}, {Wu}, {Li}, {Lei}, \& {Wu}}]{2025arXiv250110703W}
{Wang}, M., {Ma}, Y., {Li}, H., {et~al.} 2025, \bibinfo{title}{{Simulation of Binary-Single Interactions in AGN Disk I: Gas-Enhanced Binary Orbital Hardening},} arXiv e-prints, arXiv:2501.10703, \dodoi{10.48550/arXiv.2501.10703}

\bibitem[{M.~L. Waskom(2021)Waskom}]{Waskom2021}
Waskom, M.~L. 2021, \bibinfo{title}{seaborn: statistical data visualization,} Journal of Open Source Software, 6, 3021, \dodoi{10.21105/joss.03021}

\bibitem[{Y. {Yang} {et~al.}(2019{\natexlab{a}}){Yang}, {Bartos}, {Haiman}, {Kocsis}, {M{\'a}rka}, {Stone}, \& {M{\'a}rka}}]{2019ApJ...876..122Y}
{Yang}, Y., {Bartos}, I., {Haiman}, Z., {et~al.} 2019{\natexlab{a}}, \bibinfo{title}{{AGN Disks Harden the Mass Distribution of Stellar-mass Binary Black Hole Mergers},} \apj, 876, 122, \dodoi{10.3847/1538-4357/ab16e3}

\bibitem[{Y. {Yang} {et~al.}(2019{\natexlab{b}}){Yang}, {Bartos}, {Gayathri}, {Ford}, {Haiman}, {Klimenko}, {Kocsis}, {M{\'a}rka}, {M{\'a}rka}, {McKernan}, \& {O'Shaughnessy}}]{2019PhRvL.123r1101Y}
{Yang}, Y., {Bartos}, I., {Gayathri}, V., {et~al.} 2019{\natexlab{b}}, \bibinfo{title}{{Hierarchical Black Hole Mergers in Active Galactic Nuclei},} \prl, 123, 181101, \dodoi{10.1103/PhysRevLett.123.181101}

\bibitem[{W. {Yuan} {et~al.}(2019){Yuan}, {Riess}, {Macri}, {Casertano}, \& {Scolnic}}]{2019ApJ...886...61Y}
{Yuan}, W., {Riess}, A.~G., {Macri}, L.~M., {Casertano}, S., \& {Scolnic}, D.~M. 2019, \bibinfo{title}{{Consistent Calibration of the Tip of the Red Giant Branch in the Large Magellanic Cloud on the Hubble Space Telescope Photometric System and a Redetermination of the Hubble Constant},} \apj, 886, 61, \dodoi{10.3847/1538-4357/ab4bc9}

\bibitem[{F. {Zhang} {et~al.}(2019){Zhang}, {Shao}, \& {Zhu}}]{2019ApJ...877...87Z}
{Zhang}, F., {Shao}, L., \& {Zhu}, W. 2019, \bibinfo{title}{{Gravitational-wave Merging Events from the Dynamics of Stellar-mass Binary Black Holes around the Massive Black Hole in a Galactic Nucleus},} \apj, 877, 87, \dodoi{10.3847/1538-4357/ab1b28}

\bibitem[{H.-H. {Zhang} {et~al.}(2024){Zhang}, {Zhu}, \& {Yu}}]{2024arXiv240610904Z}
{Zhang}, H.-H., {Zhu}, J.-P., \& {Yu}, Y.-W. 2024, \bibinfo{title}{{Propagation of GRB Relativistic Jets in AGN Disks and Its Implication for GRB Detection},} arXiv e-prints, arXiv:2406.10904, \dodoi{10.48550/arXiv.2406.10904}

\bibitem[{Z.-H. {Zhou} {et~al.}(2023){Zhou}, {Zhu}, \& {Wang}}]{2023ApJ...951...74Z}
{Zhou}, Z.-H., {Zhu}, J.-P., \& {Wang}, K. 2023, \bibinfo{title}{{High-energy Neutrino Production from AGN Disk Transients Impacted by the Circum-disk Medium},} \apj, 951, 74, \dodoi{10.3847/1538-4357/acd380}

\bibitem[{J.-P. {Zhu}(2024){Zhu}}]{2024MNRAS.528L..88Z}
{Zhu}, J.-P. 2024, \bibinfo{title}{{High-energy neutrinos from merging stellar-mass black holes in active galactic nuclei accretion disc},} \mnras, 528, L88, \dodoi{10.1093/mnrasl/slad176}

\bibitem[{L. {Zwick} {et~al.}(2025){Zwick}, {Tak{\'a}tsy}, {Saini}, {Hendriks}, {Samsing}, {Tiede}, {Rowan}, \& {Trani}}]{2025arXiv250324084Z}
{Zwick}, L., {Tak{\'a}tsy}, J., {Saini}, P., {et~al.} 2025, \bibinfo{title}{{Environmental effects in stellar mass gravitational wave sources I: Expected fraction of signals with significant dephasing in the dynamical and AGN channels},} arXiv e-prints, arXiv:2503.24084, \dodoi{10.48550/arXiv.2503.24084}

\end{thebibliography}

\end{sloppypar}\end{document}